\documentclass[iop, numberedappendix ]{emulateapj}

\usepackage{graphicx}
\usepackage{amsmath}
\usepackage{natbib}
\usepackage{xcolor}
\usepackage[backref,breaklinks,colorlinks,citecolor=blue]{hyperref} 
\usepackage[all]{hypcap} 
\usepackage{amsmath}
\usepackage{float}

\usepackage{setspace}

\begin{document}

\title{Atmospheric variability driven by radiative cloud feedback in brown dwarfs and directly imaged extrasolar giant planets}
\author{Xianyu Tan and Adam P. Showman}
\affil{Lunar and Planetary Laboratory, University of Arizona, 1629 University Boulevard, Tucson, AZ 85721 \\  xianyut@lpl.arizona.edu}


\begin{abstract}
Growing observational evidence has suggested active meteorology in the atmospheres of brown dwarfs (BDs) and directly imaged extrasolar giant planets (EGPs). In particular, a number of surveys have shown that near-IR brightness variability is common among L and T dwarfs. Despite the likelihood from previous studies that atmospheric dynamics is the major cause of the variability, the detailed mechanism of the variability remains elusive, and we need to seek a natural, self-consistent mechanism.   Clouds are important in shaping the thermal structure and spectral properties of these atmospheres via their opacity, and we expect the same for inducing atmospheric variability. In this work, using a time-dependent one-dimensional model that incorporates a self-consistent coupling between the thermal structure, convective mixing, cloud radiative heating/cooling and condensation/evaporation of clouds, we show that radiative cloud feedback can drive spontaneous atmospheric variability in both temperature and cloud structure under conditions appropriate for BDs and directly imaged EGPs.  The typical periods of variability are one to tens of hours with typical amplitude of the variability up to hundreds of Kelvins in effective temperature. The existence of variability is robust over a wide range of parameter space, but the detailed evolution of the variability is sensitive to model parameters.   Our novel, self-consistent mechanism has important implications for the observed flux variability of BDs and directly imaged EGPs, especially for objects whose variability evolves on short timescales. It is also a promising mechanism for cloud breaking, which has been proposed to explain the L/T transition of brown dwarfs. 
\end{abstract}
\keywords{brown dwarfs - planets and satellites: gaseous planets -  planets and satellites: atmospheres - radiative transfer - methods: numerical}

\section{introduction}

Light curve variability at infrared (IR) wavelengths is common among brown dwarfs (BDs) over a wide range of spectral types  (e.g., \citealp{BJ2001, gelino2002, clarke2008, buenzli2014,radigan2014,wilson2014,metchev2015, yang2016, leggett2016,MilesPaez2017}; see a recent review by \citealp{biller2017} and \citealp{artigau2018}) as well as a handful  directly imaged extrasolar giant planets (EGPs)  \citep{biller2015, zhou2016, biller2018}.  The variability is thought to be caused by  rotational modulation of inhomogeneous surface brightness  (e.g., \citealp{radigan2012, apai2013, karalidi2016}), which is partly supported by two-dimensional surface maps of a nearby BD \citep{crossfield2014}.  Shapes of  light curves often evolve over a timescale of a few rotation periods (e.g, \citealp{artigau2009, radigan2012, biller2013, gillon2013, metchev2015, apai2017}), suggesting rapid change of the surface features.  Multi-wavelength observations display pressure-dependent  shifts of light curve shapes for BDs of certain spectral types, indicating complex vertical structures for the surface variations \citep{buenzli2012, apai2013, yang2016}.

The inhomogeneous brightness likely results from surface patchiness comprised of horizontally varying cloud and temperature structure (e.g., \citealp{radigan2012, apai2013,buenzli2015b,karalidi2016}), but the mechanisms driving the surface patchiness and controlling its evolution remain elusive. Since clouds are tracers advected by atmospheric flows, and the clouds themselves result from dynamics, the patchiness also likely has a dynamical origin.  Only a few studies of atmospheric dynamics have been conducted under conditions appropriate for BDs and directly imaged EGPs. Local hydrodynamics simulations incorporating clouds by \cite{freytag2010} and \cite{allard2012} showed that gravity waves generated by interactions between the convective interior and the stratified layer can cause  mixing that leads to small-scale cloud patchiness. However, dynamics in these local-scale models differs substantially from that at global scales. Moreover, surface patchiness at such local scales cannot by itself cause light-curve variability unless it is also accompanied by patchiness on much larger (regional-to-global) scales, and such larger-scale patchiness would necessarily be governed by very different dynamical processes.  Using dynamical equations relevant for global scales, \cite{showman&kaspi2013}  proposed that convectively excited waves could drive a global-scale circulation in the stratified atmosphere and generate horizontal patchiness in both temperature and cloud coverage, and they presented an analytic theory of how this process could work. \cite{zhang&showman2014} performed  numerical simulations of such a global circulation using an idealized one-layer shallow-water model, with the goal of determining whether the atmospheric circulation resulting from such convective perturbations would exhibit zonal banding or patchy, isotropic turbulence. Nevertheless, three-dimensional models of this process have are in their infancy. Circulation driven by latent heating associated with condensation of silicate clouds is able to generate large-scale cloud patchiness \citep{tan2017}, but latent heating alone is rather weak to be considered as a dominant driver given that horizontal temperature differences could be large in those highly variable BDs (e.g., \citealp{karalidi2016}). Using a general circulation model coupled with parameterized thermal perturbations resulting from interactions between the convective interior and the stratified atmosphere, \cite{showman2018} showed that under conditions of relatively strong forcing and weak damping, robust zonal jets and the associated meridional circulation and temperature structure are common outcomes of the dynamics. They also demonstrated that long-term (multi months to years) quasi-periodic oscillations of the equatorial zonal jets, similar to the Quasi-biennial oscillation observed in earth’s stratosphere, can be driven by the thermal perturbations. 

To date no global dynamical model of BDs explicitly implement clouds and their radiative feedback to atmospheric flows.  Opacities of  clouds have a profound impact on the thermal structure, spectral properties and thermal evolution of BDs  (see reviews by \citealp{marley2015} and \citealp{helling2014}). This should also be true in driving a vigorous global circulation and  variability in these atmospheres, as the following reasoning suggests.   Radiative-convective equilibrium  one-dimensional (1D)  models predict very different temperature-pressure (T-P) structures for models with roughly the same internal entropy but different cloud structures (see examples in, e.g., \citealp{tsuji2002, burrows2006}). This could imply that different regions of the atmosphere (e.g., cloudy and relatively cloud-free regions) exhibit significantly different T-P profiles, and the resulting horizontal temperature differences---as well as convective forcing from below---could then generate a three-dimensional (3D) atmospheric circulation that helps to control and modulate the cloud patchiness. In addition, however, local processes within a single column (representing a regional patch on a brown dwarf) can potentially control whether that local patch is cloudy and lead to variability in that cloud patch's cloud structure over time. Because clouds are subjected to gravitational sedimentation and convective mixing,  the latter being determined by the thermal structure, the changes of thermal structure should feed back onto clouds by changing the convective mixing, potentially maintaining   variability by   nonlinear self-interactions within the system. These two effects---large-scale cloud patchiness modulated by a 3D circulation and that resulting from convective mixing and cloud-radiative processes within a single atmospheric column---can operate simultaneously in an atmosphere and  interact with each other  in a highly complex manner. Although eventually one must perform global dynamical models to truly decipher the circulation driven by these effects, such models would be extremely difficult to understand and diagnose.  To build a systematic and clear understanding on the role of clouds in driving dynamics and variability,  it is crucial to study the radiative cloud feedback starting from a 1D context where the effects can be cleanly isolated, as we do here.

 In this study,  we construct a simple time-dependent 1D  atmospheric model that discards the equilibrium assumption to demonstrate the importance of radiative  cloud feedback in driving the short-term atmospheric evolution. The time dependency is crucial because the radiative cloud feedback is essentially a non-steady process. The model incorporates a self-consistent  coupling between the thermal structure, convective mixing of tracers (including entropy, condensable vapor and cloud particles), cloud radiative heating/cooling and condensation/evaporation of clouds. The model is simple in the sense that each physical component is idealized, but yet essential physical behaviors are preserved. The purposely simplified model allows us to better  understand and diagnose the underlying mechanisms, as well as a much faster computational speed to explore the mechanism over a wide range of parameter space.   

This paper is organized as follows. We introduce the numerical implementation in Section \ref{model}. We then present results, diagnosis of mechanisms and sensitivity studies in Section \ref{results}. Finally we discuss the implications in Section \ref{discussion} and conclude in Section \ref{conclusion}.

\section{Model}
\label{model}

We model the short-term evolution of a local atmospheric column of a BD or directly imaged EGP using a time-dependent 1D model, which includes  simple treatments for radiative transfer, condensation clouds, cloud gravitational settling and convective mixing. We make several assumptions: first, cloud formation is determined by chemical equilibrium, which in our simple context implies condensation obeying a Clausius-Clapeyron-type relation. Second, we assume temperature at the model's bottom boundary remains fixed during the evolution, mimicking an atmosphere attached to a convective interior whose specific entropy does not evolve over short timescales. Third, vertical transport of heat and tracers by convection is modeled as a diffusion process, and the diffusion coefficient is  determined by local temperature lapse rate as a function of pressure and time. Fourth, we neglect the effects of condensation on gas thermodynamic properties, including the release of latent heat, and the effect of the condensates on the mean molecular weight and heat capacity. Finally, we neglect advection terms. In principle, explicit vertical advection can be imposed as ascent or descent motion in 1D models. For a clear environment to understand the role of convective mixing (modeled as diffusion), we do not include explicit vertical advection in this study.  We are interested in model where clouds first condense in the convective region. 

Equations for temperature $T$, mixing ratio of condensable vapor $q_v$ (mass ratio between condensable vapor to the background dry $\rm{H_2}+He$ gas, in unit of kg/kg) and mixing ratio of condensates $q_c$ (kg/kg) as a function of height $z$  and time $t$ are as follows (a similar set of time-dependent 1D equations can be found in \citealp{smith1995}):
\begin{equation}
\frac{dT}{dt}  = -\frac{1}{c_p \rho}\frac{\partial F}{ \partial z} +  \frac{1}{c_p \rho} \frac{\partial}{\partial z}\left(c_p \rho K_{\rm{zz}} \left[\frac{\partial T}{\partial z} - \left(\frac{\partial T}{ \partial z}\right)_{\rm{ad}}\right] \right) ,
\label{eq.temp}
\end{equation}

\begin{equation}
\begin{aligned}
\frac{dq_v}{dt} = & ~ -\frac{q_v-q_s}{\tau_c} \delta  + \frac{\min{(q_s-q_v, ~q_c)}}{\tau_c} (1-\delta) \\
   & ~ +  \frac{1}{\rho} \frac{\partial}{\partial z}\left(K_{\rm{zz}} \rho \frac{\partial q_v}{\partial z}\right) + Q_{\rm{deep}},
\label{eq.qv}
\end{aligned}
\end{equation}
\begin{equation}
\begin{aligned}
\frac{dq_c}{dt} = & ~ \frac{q_v-q_s}{\tau_c} \delta - \frac{\min{(q_s-q_v, ~q_c)}}{\tau_c} (1-\delta) \\ 
 & ~ + \frac{1}{\rho} \frac{\partial (\rho \langle q_c V_s \rangle )}{\partial z} +  \frac{1}{\rho} \frac{\partial}{\partial z}\left(K_{\rm{zz}} \rho \frac{\partial q_c}{\partial z}\right),
\label{eq.qc}
\end{aligned}
\end{equation}
where $(\partial  T / \partial z)_{\rm{ad}}$ is the adiabatic temperature lapse rate, $q_s$ the local saturation vapor mixing ratio determined by a given condensation function, local temperature and pressure, $V_s$ the settling speed of particles, $K_{\rm{zz}}$ the time- and pressure-dependent vertical convective diffusion coefficients in height coordinates, $\rho$ the gas density, $\tau_c$ the conversion timescale representing source/sink of tracers due to condensation or evaporation, $c_p$ the specific heat at constant pressure, $g$ the surface gravity and $F$ the net radiative flux. The term $Q_{\rm{deep}}=-(q_v - q_{\rm{deep}})/\tau_{\rm{deep}}$ applies only at pressure regions deeper than 50 bars, relaxing local vapor $q_v$ to the deep mixing ratio $q_{\rm{deep}}$ over a characteristic timescale $\tau_{\rm{deep}}$ which is generally set to $10^3$ s, consistent with mixing timescales over a pressure scale height near the lower model boundary. Moderate deviation of $\tau_{\rm{deep}}$ from this value does not affect model results. We have performed additional experiments with  $\tau_{\rm{deep}}=10^2$ and $10^4$ s, and the variability in these sensitivity tests are almost identical to that with $\tau_{\rm{deep}}=10^3$ s.  

The first two terms on the RHS of Eq. (\ref{eq.qv}) and (\ref{eq.qc}) are sources/sinks due to condensation and evaporation, respectively, where $\delta = 1$ if vapor $q_v$ is supersaturated and  $\delta = 0$ otherwise.\footnote{The expression  $\frac{\min{(q_s-q_v, ~q_c)}}{\tau_c}$ for evaporation is a compact form of 
\begin{equation}
\left\{ 
\begin{array}{lr}
\frac{q_s-q_v}{\tau_c}  & \quad q_c \ge q_s-q_v \\
\frac{q_c}{\tau_c} & \quad q_c < q_s-q_v,
 \end{array}
 \right.
 \end{equation}
 in which the first condition implies that the vapor field is relaxed toward saturation when the amount of cloud particles is sufficient for conversion, whereas the  second condition refers to that only the existing amount of cloud particles can be converted to vapor when particles are insufficient.
 }
  The conversion timescale $\tau_c$ is assumed very short  ($\sim 10$ s, e.g., \citealp{helling2014}) compared to the cloud settling or thermal evolution timescales. We will discuss the sensitivity  to different conversion timescale. Particle settling velocity in the third term on the RHS of Eq. (\ref{eq.qc}) is calculated following Eqs. (3)-(7) in \cite{parmentier2013},\footnote{The expression of the terminal fall velocity of a spherical particle can be found in \cite{pk2012}. For detailed references of the settling velocity in the context of gas giants, the readers are referred to references in \cite{parmentier2013} or \cite{ackerman2001}.    }  and the bracket $\langle \rangle$ represents the settling flux integrated over the particle size distribution.  We implement enstatite ($\rm{MgSiO_3}$) to represent silicate cloud, which is one of the most abundant condensates in L and T dwarfs.  Assuming  solar abundance, the expression for the total gas pressure $P_T$ (in unit of bar) at which enstatite saturates  as a function of temperature is taken from  \cite{visscher2010}:
\begin{equation}
10^4 / T  = 6.26 -0.35 \log P_T .
\label{saturation}
\end{equation}
To obtain the familiar Clausius-Clapeyron-type expression for the local vapor saturation mixing ratio $q_s$, we simply use  $q_s = P_T q_{\rm{deep}}/p$ that comes with assuming all silicate vapor condense into enstatite. The deep mixing ratio  $q_{\rm{deep}}$ is about 0.0026 using the molar fraction of Mg relative to $\rm{H_2}+He$ in solar abundance from \cite{lodders2003} and assuming all silicate vapor condense into enstatite.

In our numerical implementation, equations (\ref{eq.temp}) to (\ref{eq.qc}) are solved in pressure $p$ coordinates, which can be converted from height $z$ coordinates using the hydrostatic assumption $\partial p/\partial z=-\rho g $.  We integrate the system  forward with time using the third-order Adams-Bashforth scheme \citep{durran1991} which has been frequently used in atmospheric dynamical models.  The value of a prognostic variable, $T$ for example, at time $t+\Delta t$ is marched forward by
\begin{equation}
\begin{aligned}
T(t+\Delta t) =  & T(t) + \frac{\Delta t}{12}  [23 \frac{dT}{dt}(t)   \\
 & - 16\frac{dT}{dt}(t-\Delta t) + 5\frac{dT}{dt}(t-2\Delta t) ].
\end{aligned}
\end{equation}
At the initialization, we first integrate the system two time steps using a simple Euler scheme, then switch to the third-order Adam-Bashforth scheme.  The time step is chosen small enough (0.1 s for simulations with nominal vertical resolution) to guarantee stability and convergence. Our initial condition includes a cloud-free radiative-convective equilibrium profile, solar abundance of silicate vapor below condensation level and no clouds. After sufficient integration time, the system reaches a statistically equilibrium state which is insensitive to initial conditions.\footnote{The statistical equilibrium is not steady. As will be shown in next section, there are large temporal fluctuations around what amounts to a mean ``climate''. We have tested the models with vastly different initial conditions. When reaching statistical equilibrium, the mean climate of the system---including the amplitude and period of the variability---is independent of initial conditions.} The computation domain ranges from $10^{-3}$ to 100 bars that is discretized into 100 layers. We have tested additional models with different vertical resolutions of 50, 200 and 300 layers (shown in the Appendix \ref{ch. verticalresolution}). Results show good agreement for resolution of 100 layers or more. For computational efficiency, we use 100 layers for models presented in this work. We emphasis that our model does not implement external {\it{ad hoc}} forcing like those in \cite{zhang&showman2014} and \cite{robinson2014}, and the atmospheric activity shown below arise spontaneously from self-interactions of the system.

\subsection{Cloud Size Distribution}
\label{ch.clouddistribution}
We assume a constant cloud particle number  per dry air mass  $\mathcal{N}_c$ (in unit of $\rm{kg^{-1}}$) throughout the atmospheric column, then use this number to determine local cloud properties---such as the time-varying, pressure-dependent mean particle size---given the time- and pressure-dependent amount of condensate.   In our nominal models, we prescribe a log-normal  particle size distribution, which has been widely used to parameterize clouds in BD  atmospheres (e.g., \citealp{ackerman2001, barman2011a, morley2012}): 
\begin{equation}
n(r) = \frac{\mathcal{N}_c}{\sqrt{2\pi} \sigma r} \exp \left( - \frac{\left[ \ln(r/r_0)\right]^2}{2\sigma^2} \right),
\end{equation}
where $r$ is the particle radius,  $n(r)=d\mathcal{N}_c/dr$ the number density distribution, $r_0$ is a reference radius, and $\sigma$ is a nondimensionalized constant that measures the width of the distribution. The number density of particles peaks at a radius slightly less than $r_0$, and exponentially decreases with radius for radii smaller or larger than the radius corresponding to this peak. This function aims at capturing a possibly broad spread of particle size distribution (see a detailed discussion in \citealp{ackerman2001}). However, the parameter $\sigma$ controlling the width of size distribution  is unconstrained. Here for the nominal models, we empirically assume $\sigma=1$, and will test the sensitivity to varying $\sigma$. With the cloud mass-mixing ratio $q_c$ obtained from Eq. (\ref{eq.qc}), and given the specified values of $\sigma$ and $\mathcal{N}_c$, the particle size distribution can be determined by calculating the reference radius $r_0$. This is done by solving the integral $q_c = \frac{4}{3}\pi \rho_c \int_{0}^{\infty} r^3 n(r) dr = \frac{4}{3}\pi \rho_c \mathcal{N}_c r_0^3  \exp\left(\frac{9}{2}\sigma^2\right)$, which is simply the definition of the cloud mass mixing ratio.  Practically, when summing over the particle size distribution for calculation of cloud opacity and gravitational settling flux, we discard particles smaller than  a minimal  radius $r_{\rm{min}}$ and a maximum radius $r_{\rm{max}}$, the latter of which is set to 100 $\mu $m.  We choose $0.01 ~ \rm{\mu m}$ for  $r_{\rm{min}}$ following \cite{tsuji2002} who estimated   that in typical brown dwarf conditions, $0.01 ~ \rm{\mu m}$ is roughly the criteria below which the cloud particle may not be stable against surface tension. In fact, our numerical results are not sensitive to the choice of $r_{\rm{min}}$ as long as it is sufficiently small, because opacity contributions from particles smaller than $\sim$$0.01\,\mu$m are negligible in determining the heating/cooling rates that drive the evolution.

We also have implemented an exponential particle size distribution which has been widely used to describe precipitation in Earth's atmosphere (e.g., Chapter 2, \citealp{straka2009}):
\begin{equation}
 n(r) = \frac{\mathcal{N}_c}{r_0} \exp \left( - \frac{r}{r_0} \right),
 \label{eq.expcloud}
\end{equation} 
to test the sensitivity to different assumed size distribution. The detailed evolution of atmospheric variability differs with different size distribution function, but the mechanism---and qualitative behavior---remain the same. For conciseness, we will not present results of the exponential size distribution in the main text.  However, Appendix \ref{ch.expcloud} presents some tests demonstrating that models with both size-distribution functions exhibit qualitatively similar behavior.

\subsection{Radiative Transfer}
We model the atmospheric radiative transfer using a plane-parallel, two-stream approximation. We focus on a grey atmosphere---with a single broad thermal band---for simplicity and computational efficiency. The radiative transfer equations in an absorbing, emitting and multiple-scattering atmosphere with the $\delta$-function adjustment for scattering are solved using an efficient numeral package TWOSTR (\citealp{kst1995}, wherein the radiative equations solved in our model can be found). \cite{komacek2017} present tests quantifying the model behavior relative to analytically known solutions for simple cases; these tests demonstrate that our numerical implementation agrees extremely well with known analytic solutions. The background model atmosphere uses a frequency-averaged gas opacity, the Rosseland-mean opacity $\kappa_{\rm{R, g}}$, in all pressures from \cite{freedman2014} with solar composition. Since we are mostly interested in the cloud condensation region where it is optically thick, the Rosseland-mean opacity gives a good estimation of radiation flux in this limit. In the upper atmosphere where it is optically thin, there is no good choice {\it a priori} for a single opacity in the grey approximation. 

 Cloud particles interact with radiation via absorption and scattering, which are parameterized by the extinction coefficient $Q_{\rm{ext}}$, scattering coefficient $Q_{\rm{scat}}$ and asymmetry parameter $\tilde{g}$.  To be consistent with the background Rosseland-mean gaseous opacity, the total cloud extinction opacity $\kappa_{\rm{R, ext}}$ is averaged over wavelength using the Rosseland-mean definition:
\begin{equation}
\frac{1}{\kappa_{\rm{R, ext}}} = \frac{\int_0^{\infty} \frac{1}{\kappa_{\rm{ext}}(\lambda)} \frac{dB_{\lambda}}{dT} d\lambda}{\int_0^{\infty}  \frac{dB_{\lambda}}{dT} d\lambda},
\end{equation}
where $B_{\lambda}$ is the Planck function and $\kappa_{\rm{ext}}(\lambda) = \int_{r_{\rm{min}}}^{\infty} n(r)\pi r^2 Q_{\rm{ext}}(r,\lambda) dr $ is the total cloud opacity at $\lambda$ summing over all particle sizes. The cloud scattering opacity $\kappa_{\rm{R,scat}}$ and $\tilde{g}$ are defined likewise. Assuming spherical particles, the coefficients $Q_{\rm{ext}}(r, \lambda)$, $Q_{\rm{scat}}(r, \lambda)$ and $\tilde{g}(r, \lambda)$ are pre-calculated with Mie theory using the numerical package written by   \cite{schafer2012} with the refractive index of enstatite obtained from \cite{jager2003}. The total opacity is simply the sum of gas and cloud opacity $\kappa = \kappa_{\rm{R, g}} + \kappa_{\rm{R, ext}}$.  Note that although we adopt the specific optical properties of enstatite particles for our nominal models, our qualitative results remain robust over a wide range of cloud optical properties.  In addition to our nominal models, we performed a range of simulations varying the scattering coefficient, extinction coefficient, and asymmetry parameter over plausible ranges; these simulations demonstrate overall behavior (including temporal variability) that qualitatively resembles those of our nominal models.

\subsection{Diffusion}
Convective mixing for entropy and tracers is modeled as diffusion processes. The time- and pressure-dependent vertical diffusion coefficient $K_{\rm{zz}}$  in height coordinate depends on the temperature lapse rate. We adopt the classical expression of the diffusion coefficient based on the mixing length theory \citep{gierasch1968}: 
\begin{equation}
K_{\rm{zz}} = \left\{
\begin{array}{lr}  
0, \quad \quad \frac{\partial \ln T}{\partial \ln p} - \left(\frac{\partial \ln T}{\partial \ln p} \right)_{\rm{ad}} \leq 0 \\
\\
l^2 \frac{g}{\sqrt{RT}}\sqrt{\frac{\partial \ln T}{\partial \ln p} - \left(\frac{\partial \ln T}{\partial \ln p} \right)_{\rm{ad}}}, \quad \quad \rm{else}
\end{array}
\right.
\label{eq.kzz}
\end{equation}
where $l$ is the mixing length, which we set equal to the local pressure scale height $H_p = \frac{RT}{g}$. By definition, the convective diffusion coefficient is only effective in the convective region where $\frac{\partial \ln T}{\partial \ln p} - \left(\frac{\partial \ln T}{\partial \ln p} \right)_{\rm{ad}} > 0$, so outside the convective region $K_{\rm{zz}}$ is set to zero in the nominal setup.  However, small-scale eddies  or large-scale flows could provide transport of tracers in the stratified regions. Because quantifying the mixing in the stratified atmosphere is difficult as the atmospheric conditions are unknown, in some experiments we impose a minimal value of  $K_{\rm{zz}}$  as a background value to the diffusion coefficient for tracers (cloud and vapor)  to crudely represent mixing in the stratified layers, and will discuss the results in the sensitivity studies section.

\begin{figure*}      
\epsscale{1.}      
\plotone{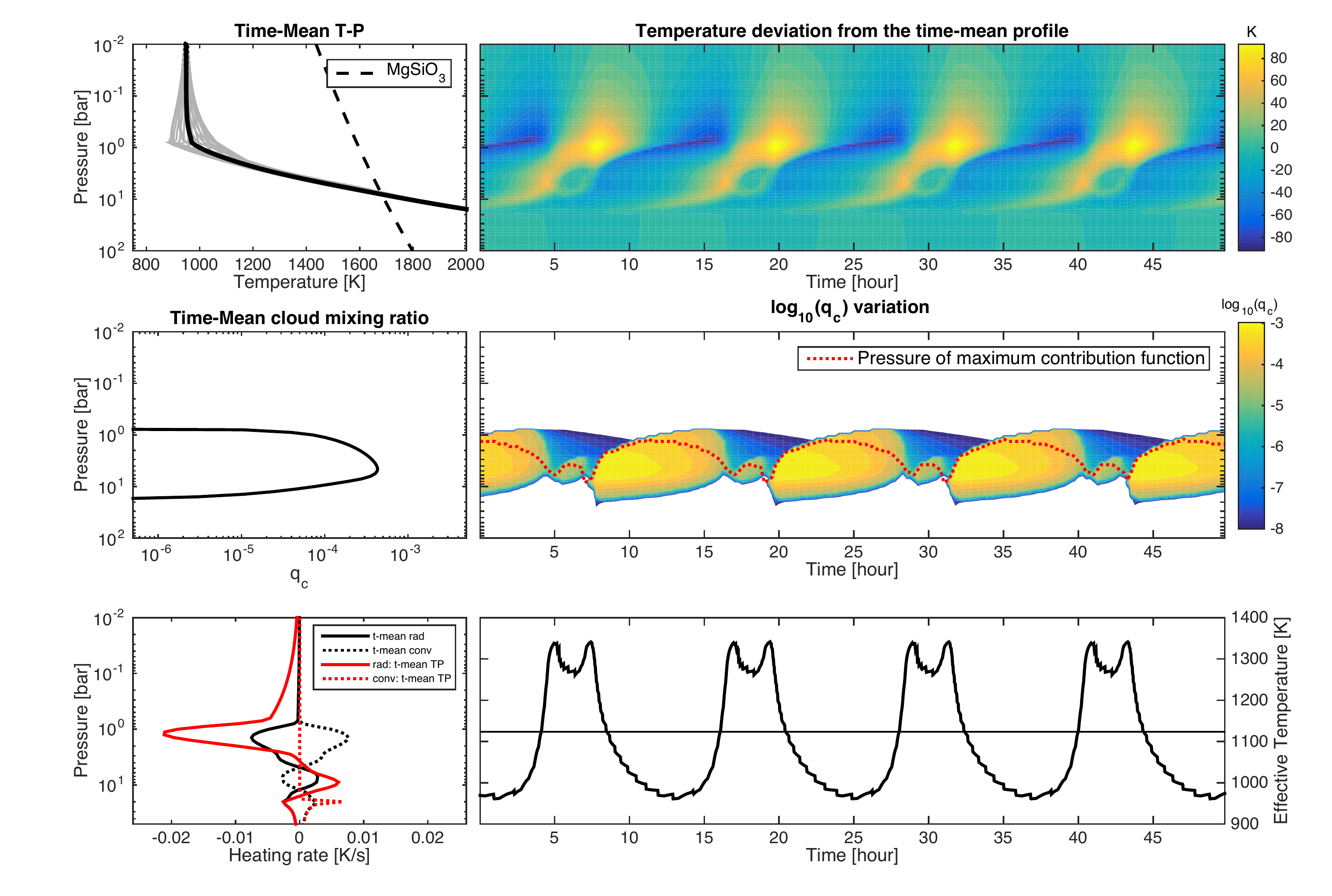}
\caption{Evolution of temperature, cloud mixing ratio and outgoing thermal flux from the nominal model with a log-normal cloud size distribution with $\sigma=1$,  cloud number per mass $\mathcal{N}_c=5\times10^8 ~ \rm{kg^{-1}}$, bottom temperature 3400 K at 100 bars and surface gravity $g=10^3 ~\rm{m/s^2}$. Left columns show time-mean temperature in the upper panel (the dashed line represents enstatite condensation curve, and the grey lines represent the envelope of the variation range of the temperature-pressure profile), and cloud mixing ratio in the middle panel. The lower left panel contains several heating rate profiles: the black solid line represents time-mean radiative heating rate; the black dotted line is time-mean convective heating rate; the red solid line and red dotted line are the radiative and convective heating rates calculated using the time-mean temperature, cloud-pressure profiles, respectively.  The upper right panel is a color plot of temperature deviations from time-mean values as a function of time and pressure, and likewise the middle right panel shows contours for the cloud mixing ratio in logarithmic scale. The dotted red line is the pressure at which the contribution function is at a maximum, which is where the majority of flux escape from the atmosphere.  The lower right panel is the corresponding outgoing thermal flux as a function of time plotted as an effective temperature, and the horizontal flat line is the time-mean effective temperature.}
\label{timeplot1}
\end{figure*} 

\section{Results}
\label{results}

\subsection{A Nominal Case}
\label{nominal}
We start by describing a nominal model with a log-normal cloud size distribution with $\sigma=1$, cloud number per mass $\mathcal{N}_c=5\times10^8 ~ \rm{kg^{-1}}$, bottom temperature 3400 K at 100 bars and surface gravity $g=10^3 ~\rm{m~s^{-2}}$. If there were no clouds, the equilibrium solution  would have an effective temperature $\sim1380$ K.  The time-averaged T-P profile is shown in the upper left column of Figure \ref{timeplot1}, and the time-averaged  cloud mixing ratio is shown in the middle left column. The intersection between the condensation curve and the mean T-P profile indicates that the predicted condensation level is  at about 8 bars. The time-mean cloud layer extends vertically across more than two pressure scale heights. 
The cloud structure undergoes significant variation over short timescales while the temperature structure varies moderately.  The grey lines in the upper left panel of Figure \ref{timeplot1} represent an envelope of  T-P variations, and  the right column shows (from top to bottom) the temperature deviations relative to the time-averaged profile, the cloud mixing ratio as a function of time and pressure, and the outgoing thermal flux (expressed as an effective temperature), respectively, all as a function of time. The evolution of these quantities shows fairly regular oscillation patterns with a period of about 12 hours. The temperature variation on isobars reaches a maximum of about 180 K at around 0.8 bar,  and gradually decrease above and below this level. The cloud mixing ratio and layer thickness vary significantly during the evolution, ranging from a thick cloud layer that extends more than two pressure scale heights to an almost cloud-free atmosphere.  The time-averaged outgoing thermal flux is $\sim 1125$ K in terms of effective temperature, much lower than that of the cloud-free model; this difference resuls from the higher and thus cooler emission level associated with the cloud top.\footnote{Here the cloud top is loosely defined as the level below which the cloud opacity rapidly exceeds the background gaseous opacity.}  The amplitude of the effective temperature variation exceeds 350 K, much larger than the actual temperature variation on isobars. The outgoing flux variation  is mainly caused by altitude variation of the cloud top and thus the emission-level temperature variation, though the actual temperature variations (on isobars) also contribute positively. This is demonstrated by showing that  the pressure level at which the  contribution function reaches a maximum (the dotted line in the middle panel of Figure \ref{timeplot1})  closely follows the evolution of the cloud top. 

The system is in a statistically equilibrium state over long timescales, meaning that the time-averaged radiative flux divergence is balanced by convective flux divergence, and the tracer settling flux is balanced by convective mixing flux. This is illustrated   in the lower left panel of Figure \ref{timeplot1}, where the time-averaged radiative and convective heating rates (the black solid and dotted lines, respectively) are equal in magnitude but of opposite sign. However, this equilibrium state cannot be characterized using the time-averaged T-P and cloud profiles due to the variability and the nonlinearity between temperature and flux.   The radiative and convective heating rates of the time-averaged temperature and cloud profiles are also plotted in  the lower left panel of  Figure \ref{timeplot1} as red lines.  The significant nonzero net heating rate suggests that this single set of time-averaged profiles is not in equilibrium. In addition, the outgoing thermal flux of the time-averaged profiles is not equal to the time-averaged outgoing thermal flux.  This is always the case for  models showing variability. This makes sense because the thermal and tracer profiles are imbalanced at any instant, and the time averages of them  are therefore not necessarily in equilibrium. Our results suggest that a single set of T-P profiles cannot   represent a statistical equilibrium of an atmosphere with vigorous cloud formation.

There are small-scale jumps in the flux curve shown in Figure \ref{timeplot1}, which are caused by the rapid changes of the cloud-top pressure when a new convectively unstable layer formed above the cloud top.   The period and amplitude of the jumps in flux depend on the numerical resolution of the pressure grid. As shown in our resolution test  in Appendix \ref{ch. verticalresolution}, higher resolution results in a smoother flux curve, but the overall quantitative variability remains almost the same. We emphasis that these jumps are not numerical errors, but physical behaviors resulting from finite numerical resolution. 

\begin{figure*}      
\epsscale{1.2}      
\plotone{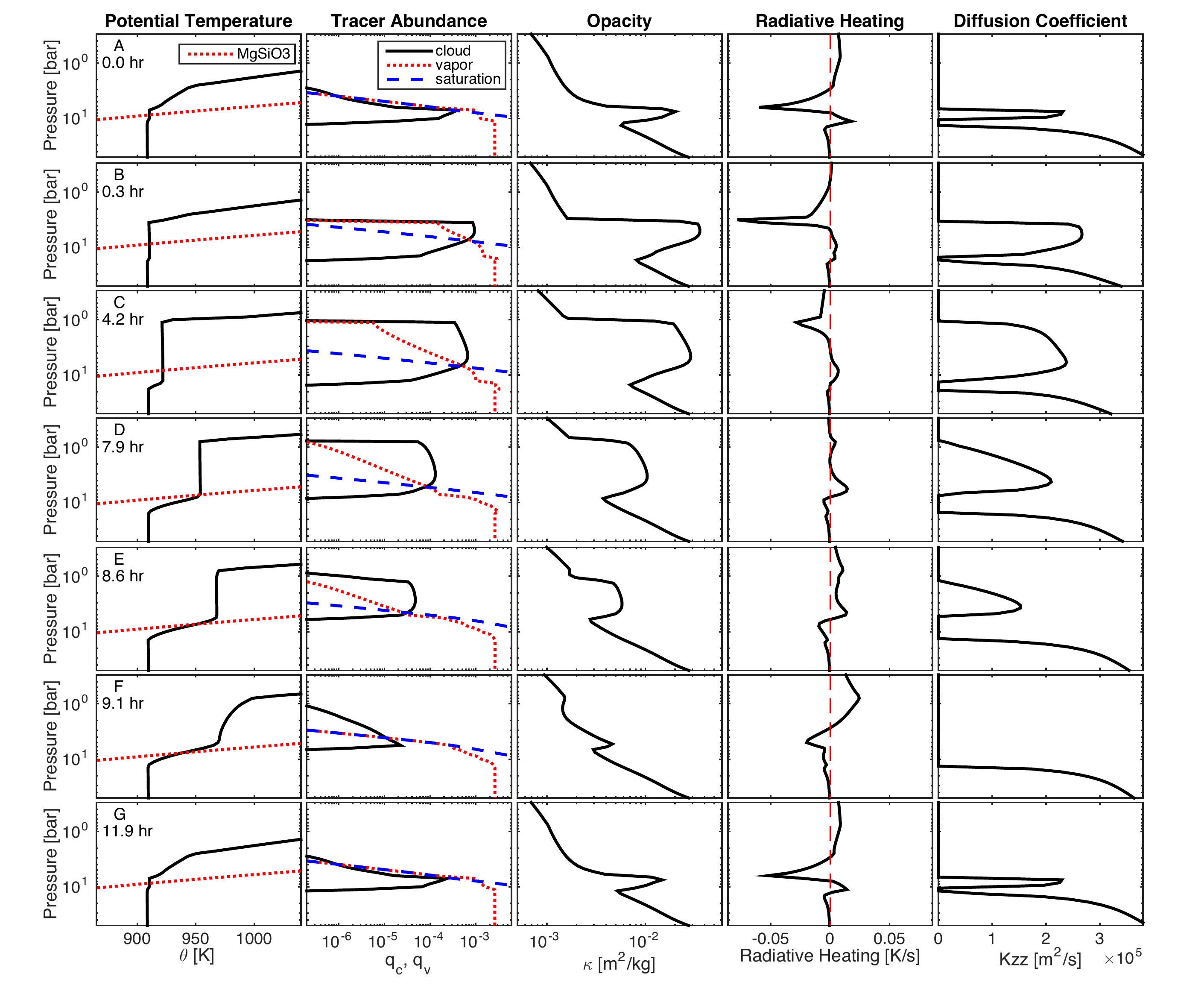}
\caption{Here show several diagnostic quantities that are used to illustrate the detailed evolution and mechanism of the variability. The sequence is shown from a complete cycle of the nominal case shown in Figure \ref{timeplot1}.  Each row represents a snapshot of the atmospheric state at a given time with time increasing downward as marked at the beginning of each row. The first column is the potential temperature $\theta$, and the dashed lines show the enstatite ($\rm{MgSiO_3}$) condensation curve in $\theta$-pressure space. The second column contains cloud mixing ratio $q_c$ in solid lines, vapor mixing ratio $q_v$ in dotted lines and saturation vapor mixing ratio $q_s$ in dashed lines. The third column is the total opacity $\kappa$ and forth column is radiative heating/cooling rate. Finally the fifth column shows the vertical diffusion coefficient $K_{\rm{zz}}$. 
}
\label{mechanism}
\end{figure*} 

Next we show various diagnostic quantities in an evolution cycle of the system, which helps to clarify the mechanism by which the variability is driven.  Figure \ref{mechanism} contains the quantities at  different time snapshots.  The potential temperature in the first column is defined as $\theta = T(p_0/p)^{R/c_p} $, referring to the temperature an air parcel would have if it is adiabatically compressed or expensed to a reference pressure $p_0$, and $R$ the specific gas constant.  We choose $\theta$ instead of temperature to represent thermal structure because  convective regions can be easily recognized as having a nearly constant $\theta$ with pressure. Tracer abundances, including vapor mixing ratio $q_v$, saturated vapor mixing ratio $q_{v,s}$ and cloud mixing ratio $q_c$ are shown in the second column; the total grey thermal opacity is shown in the third column; the radiative heating/cooling rate is shown in the fourth column and finally the diffusion coefficient $K_{\rm{zz}}$ is shown in the fifth column.

Comparisons between the potential temperature $\theta$, cloud mixing ratio $q_c$ and diffusion coefficient $K_{\rm{zz}}$ profiles suggest that the cloud layer always corresponds  to a secondary detached convective layer. In other words, the formation of clouds generates a stratification below the cloud base where otherwise would be a convective region. The detached convective zone associated with cloud formation has also been revealed by many  studies (e.g., \citealp{tsuji2002, burrows2006}). Large cloud opacity creates an opacity inversion near the cloud base (see the opacity structure in the third column), which leads to heating below the cloud base, stratifying the nearby layers. Meanwhile, strong radiative cooling occurs at the cloud top due to the sharp vertical opacity gradient caused by the sharp cloud top. The combination of cooling at the top and heating at the bottom maintains a convective instability within the cloud layer which is responsible for the secondary convective layer.  In turn, the secondary convective layer provides strong mixing  that  can almost well mix the bulk particle mixing ratio within the layer, maintaining the cloud layer against rapid gravitational settling and creating a sharp cloud top. The mixing efficiency can be estimated by comparing two timescales: the diffusion timescale over a scale height $H_p^2/K_{\rm{zz}}$ and settling timescale $H_p/v_{\rm{set}}$, where $v_{\rm{set}}$ is the settling velocity in height coordinates (which can be found in, e.g., Eq. (3) in \citealp{parmentier2013}). For example, based on the model solution, the typical diffusion coefficient in the cloud layer is  $\sim 2\times 10^5 ~\rm{m^2 s^{-1}}$; this value implies a diffusion timescale similar to settling timescale for particles of $\sim$ 40 $\mu $m. This implies that particles smaller than $\sim$ 40 $\mu $m can easily  be lofted by convective mixing. In this particular simulation the particle size distribution peaks at around 0.4 $\mu $m when the cloud layer is thick, and so the majority of particle settling flux can be roughly balanced by the mixing within the cloud layer.    Above the cloud top, the  atmosphere is stratified; otherwise convection  would quickly mix clouds upward until reaching the stratified region. There, $K_{\rm{zz}}$ is zero -- or for practical purposes -- much smaller than the  $K_{\rm{zz}}$ in the convective region, so is unable to mix much cloud,  therefore creating a sharp cloud top. 

Results show high supersaturation of condensable vapor in the secondary convective zone -- the vapor abundance is much higher than the equilibrium saturation vapor abundance determined by Eq. (\ref{saturation}). This can be seen  in the second column of Figure \ref{mechanism}. With a finite conversion timescale $\tau_c$ between vapor and cloud, the high supersaturation is caused by the strong convective mixing.  Because of the abundant vapor at altitudes slightly above the condensation level, the source of clouds there is actually from condensation, with a nearly balanced sink from convective mixing that distributes  clouds toward higher altitudes and below the condensation level. To quantify the degree of supersaturation as a function of  $\tau_c$ and   $K_{\rm{zz}}$,  one can compare the lapse rate of equilibrium vapor mixing ratio $d\ln q_s/d\ln p$ (which is determined only by temperature) to that of the actual vapor mixing ratio $d\ln q_v/d\ln p$ (which is determined by both microphysical and mixing timescales). Using Eq. (\ref{saturation}), relation $q_s = P_T q_{\rm{deep}}/p$ and assuming a nearly adiabatic T-P profile ($d\ln T/d\ln p \sim R/c_p$), one can get $d\ln q_s/d\ln p = \frac{R}{c_p} \frac{65788}{T} -1$. The lapse rate of the actual vapor mixing ratio can be estimated by assuming  a quasi balance between condensation and convective mixing in the pressure-coordinate version of Eq. (\ref{eq.qv}): $\frac{q}{\tau_c} \sim \frac{d}{d p}\left(K_{\rm{zz}} (\rho g)^2 \frac{d q_v}{d p}\right) $ where we assume $q_s \ll q$. Treating $K_{\rm{zz}}$ and $d\ln q_v/d\ln p$ as  nearly constants on the local scale, we have $d\ln q_v/d\ln p \sim \sqrt{\frac{R^2T^2}{\tau_c K_{\rm{zz}} g^2} + \frac{1}{4} } - \frac{1}{2}$. 
Thus, the ratio of lapse rates is
\begin{equation}
\frac{d\ln q_s}{d\ln p} /\frac{d\ln q_v}{d\ln p} \sim \frac{\frac{R}{c_p} \frac{65788}{T} -1}{ \sqrt{\frac{R^2T^2}{\tau_c K_{\rm{zz}} g^2} + \frac{1}{4} } - \frac{1}{2} }.
\label{eq.supersaturation}
\end{equation}
Assuming $T \sim 1300$ K, $K_{\rm{zz}} \sim 2\times 10^5 ~\rm{m^2 s^{-1}}$ and $\tau_c=10$ s, one can get the ratio around 4.5, meaning that the supersaturation $q_v/q_s$ could reach the order of 100 only one pressure scale height above the condensation level,  roughly consistent with the model results. For the same reason, some fraction of clouds can be mixed well below the condensation level even with a very short evaporation timescale due to  efficient convective mixing.  

A few features are interesting. First, temperature variations exhibit a pressure-dependent shift as seen from the tilted patterns in upper right panel of Figure \ref{timeplot1}, and the maximum phase difference is between the cloud top and base. This  shift  is a result of transition between convection and stratification in the cloud forming region shown in Figure \ref{mechanism}. Second,  thick clouds usually correlate to a convective and cooler profile, and thin clouds correlate to a stratified and warmer profile. This is because the secondary convective layer and the cloud layer are coupled and coevolve in time. When clouds are present, the atmospheric structure is forced to be nearly adiabatic which is cool above the condensation level. When cloud dissipates, the region where there were clouds is stratified and thus warmer. Finally, the instantaneous cloud-base pressure varies substantially during the evolution, and differs from the prediction of  the intersection of the condensation curve described by Eq. (\ref{saturation}) and the time-mean T-P profile. This is not surprising because the actual condensation level is determined by the instantaneous vapor and temperature profiles, both of which vary during the evolution. 

\subsection{Mechanism of the Variability}
\label{ch.mechanism}

The evolution of the system is mainly comprised of the formation and dissipation of clouds and the corresponding transition between convective and stratified thermal profile, as shown in both Figure \ref{timeplot1} and \ref{mechanism}.  In between the cycle, the thickness of the cloud layer as well as the altitude of the cloud top gradually grow and then decay. Here we discuss the mechanisms that control the behavior and maintain the variability.

After the cloud layer forms, it dissipates primarily by particle settling through the cloud base. The stratification right below the cloud base plays a key role to suppress efficient upward transport of both condensable vapor and clouds through the cloud base.    Thus, there is no source to balance the loss of the total amount of tracer including clouds and vapor in the cloud forming region, and inevitably the cloud layer will eventually decay.   The secondary convective layer that coexists with the cloud layer cannot prevent the net sinking by particle settling unless it can penetrate below the cloud base and mix up vapors. However, this is not physically feasible due to the strong heating rate at the cloud base which  stratifies regions below the cloud base as shown in Figure \ref{mechanism}.  One could wonder whether this result could be an artifact of using diffusion  to parameterize convective transport. In reality, convection has upward and downward motions. In regions of strong updrafts cloud settling may be prevented, but in regions of downdrafts  falling out of clouds is inevitable and so there will always be net rainout through the cloud base. 

After the cloud layer dissipates, the atmosphere will cool off due to the larger outgoing radiative flux. Relative to the temperature at deeper level which is approximately constant in time, the cooling of the upper layers lessens the stratification of the profile (i.e., makes $d\ln T/d\ln p$ larger), and eventually promotes upward convective mixing of deep vapor. In some situations the cooling also triggers supersaturation of substantial amount of vapor. In both cases, a new cloud deck can form and the cycle repeats. Of course this simplified picture does not describe the full details such as the small sub-cycle seen in Figure \ref{timeplot1}, but demonstrates the qualitative governing mechanism for the cloud cycle. 

\begin{figure*}
\epsscale{0.8}
\plotone{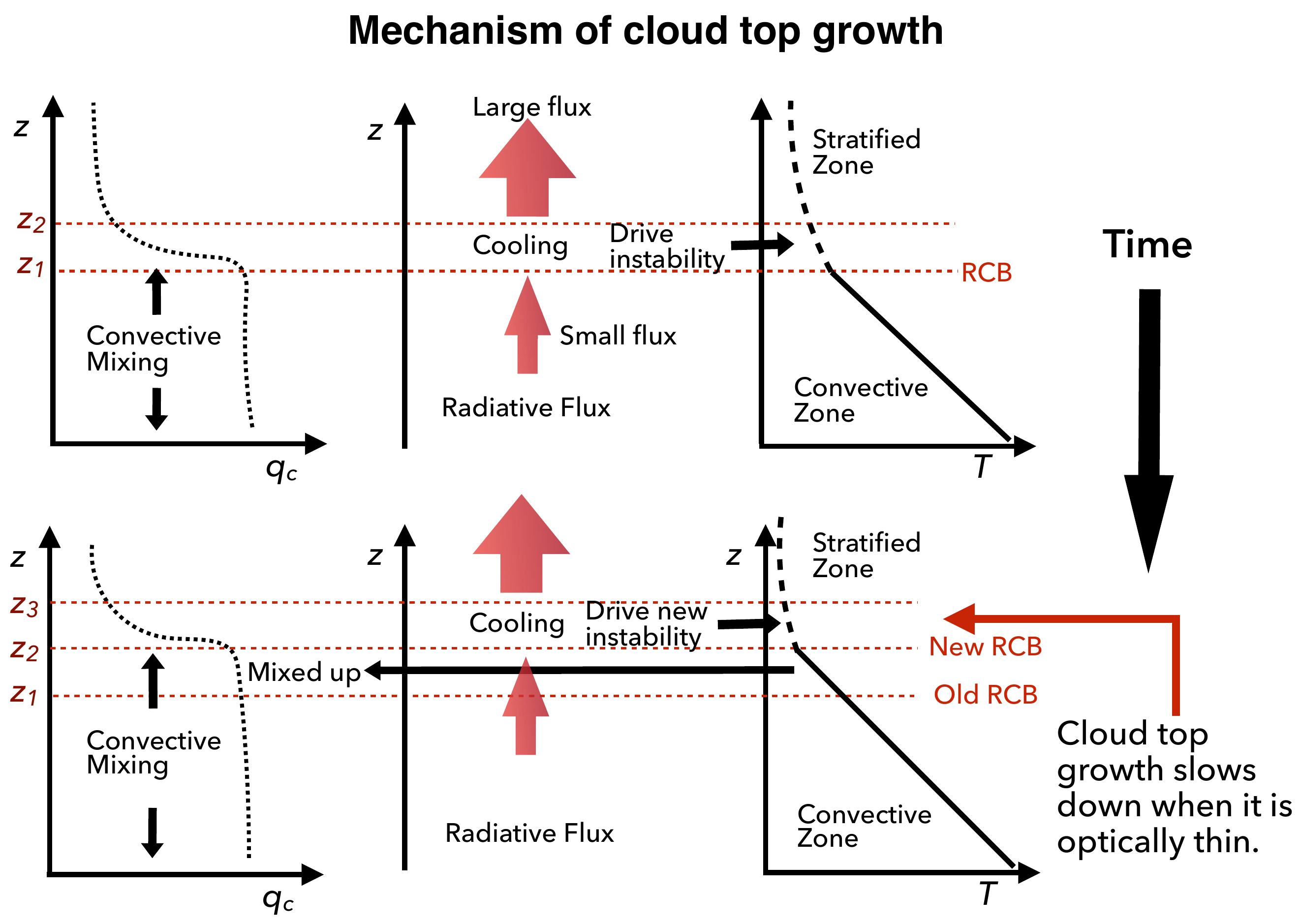}
\caption{Illustration of the mechanism for cloud top growth -- the generation of convective instability by the vertically sharp cloud top.  Suppose that initially the cloud top as well as the top of the secondary convective zone lies at altitude $z_1$, and there exists a transition to a nearly cloud-free altitude $z_2$ which is slightly above $z_1$.  Convective mixing is strong enough to almost well mix clouds up to $z_1$, above which mixing is too small to maintain much cloud.  This results in a sharp transition of cloud mixing ratio and thus a large opacity gradient between $z_1$ and $z_2$. This configuration drives a strong cooling for the layer between $z_1$ and $z_2$.  After some time the temperature at $z_2$ is sufficiently cool and the layer between $z_1$ and $z_2$ becomes convectively unstable, and clouds are then mixed up to the higher altitude $z_2$. Therefore, a new sharp cloud top forms  between $z_2$ and a higher altitude $z_3$. This procedure repeats, and the cloud top is able to extend to higher altitudes over time. The ascend of the cloud top  can be slowed down or terminated either when the cloud top reaches sufficiently high in altitude where it is optically thin or when the cloud mixing ratio is sufficiently small.
}
\label{cloudtop}
\end{figure*}

Within a cloud cycle, thickness of a cloud layer and the altitude of the cloud top grow, which is responsible for the decrease of outgoing thermal flux shown in Figure \ref{timeplot1}. This requires that the secondary convective zone continuously extends to higher and higher altitude over time. The mechanism uplifting the top of the secondary convective zone is the continuous generation of convective instability in the layer just above the cloud top due to the large vertical gradient of cloud mixing ratio near the cloud top. We depict the process using a cartoon plot in Figure \ref{cloudtop}. Suppose that initially the cloud top as well as the top of the secondary convective zone lies at altitude $z_1$, and there exists a transition to a nearly cloud-free altitude $z_2$ which is slightly above $z_1$.  Convective mixing is strong enough to almost well mix clouds up to $z_1$, above which mixing is too small to maintain much cloud.  This results in a sharp transition of cloud mixing ratio and thus a large opacity gradient between $z_1$ and $z_2$. This configuration drives a strong cooling for the layer between $z_1$ and $z_2$ by the fact that radiative flux is higher above $z_2$ than below $z_1$ (see the large gradients and  a strong cooling just above the cloud top in Figure \ref{mechanism}).\footnote{In the presence of large opacity gradient that increases with depth, the atmospheric profile is usually convectively unstable if it were in radiative equilibrium (see a discussion in, e.g.,  \citealp{rauscher2012}). Larger opacity gradient leads to a more unstable equilibrium profile. In the non-equilibrium atmosphere with strong opacity gradient like the layer near the cloud top in our case,  radiation tends to drive the atmosphere towards a convectively unstable equilibrium by a strong top cooling.   }  After some time the temperature at $z_2$ is sufficiently cool and the layer between $z_1$ and $z_2$ becomes convectively unstable, and clouds are then mixed up to the higher altitude $z_2$. Therefore, a new sharp cloud top forms  between $z_2$ and a higher altitude $z_3$. This procedure repeats, and the cloud top is able to extend to higher altitudes over time. Another essential angle to understand this mechanism is that, as the cloud-top altitude rises, the thermal radiation to space decreases, and this cools off the entire temperature profile of the stratified, cloud-free atmosphere that is above the cloud. Because the opacity structure has a large vertical gradient near the cloud top,   this cooling tends to drive a large temperature lapse rate just above the cloud top and thus causes the instability to extend slightly higher in altitude. This then  cools the above-cloud temperature profile even more, causing the cloud top to extend to even higher altitude.

The ascend of the cloud top  can be slowed down or terminated by two processes. First, when the cloud top reaches sufficiently high in altitude, the atmosphere  becomes too optically thin to drive an instability, and therefore the growth of the cloud top ceases.\footnote{In radiative equilibrium, the atmosphere is stratified in the optically thin limit even in the presence of a strong opacity gradient. In this limit the cooling that drives the atmosphere towards equilibrium cannot generate convective instability. } Second, when the cloud mixing ratio is significantly reduced due to the aforementioned dissipation mechanism, the opacity gradient is reduced and so is the top cooling. For example, row D in Figure \ref{mechanism} shows that the cloud top is at optical depth $\ll 1$ and there is no top cooling; row E shows that when cloud mixing ratio is reduced there is no top cooling. Clouds start dissipating when the cloud mixing ratio significantly decreases and thus become unable to maintain sufficient convective mixing to balance the settling. Essentially, this is a runaway process that kills off the cloud -- the lessening of the cloud opacity lessens the opacity discontinuity at the cloud top, which lessens the cooling spike at the cloud top, and therefore helps to inhibit the convective instability, all of which further acts to suppress upward mixing of cloud material, allowing the cloud to die by particle settling in a stagnant environment.  Rainout starts from low pressure where the settling velocity is higher than at high pressure (see the settling velocity as a function of pressure in, e.g., \citealp{parmentier2013}).  Above levels where clouds dissipate, air is warmed up by the larger upwelling radiative flux, thus stratifying those levels.

The spontaneous and continuous variability of the system is therefore maintained by the persistent imbalances. Because of the imbalanced net cloud settling, the cloud layer will continuously dissipate until it is replenished by a new cloud deck.  Also because the cloud layer and the secondary convective zone is tightly coupled, the evolution of clouds forces the evolution of the thermal structure. The imbalance of clouds is likely inevitable because the secondary convective zone cannot extend below the cloud base. Thus, the system shown in Figure \ref{timeplot1} and \ref{mechanism} is intrinsically variable.

\begin{figure}      
\epsscale{1.3}      
\plotone{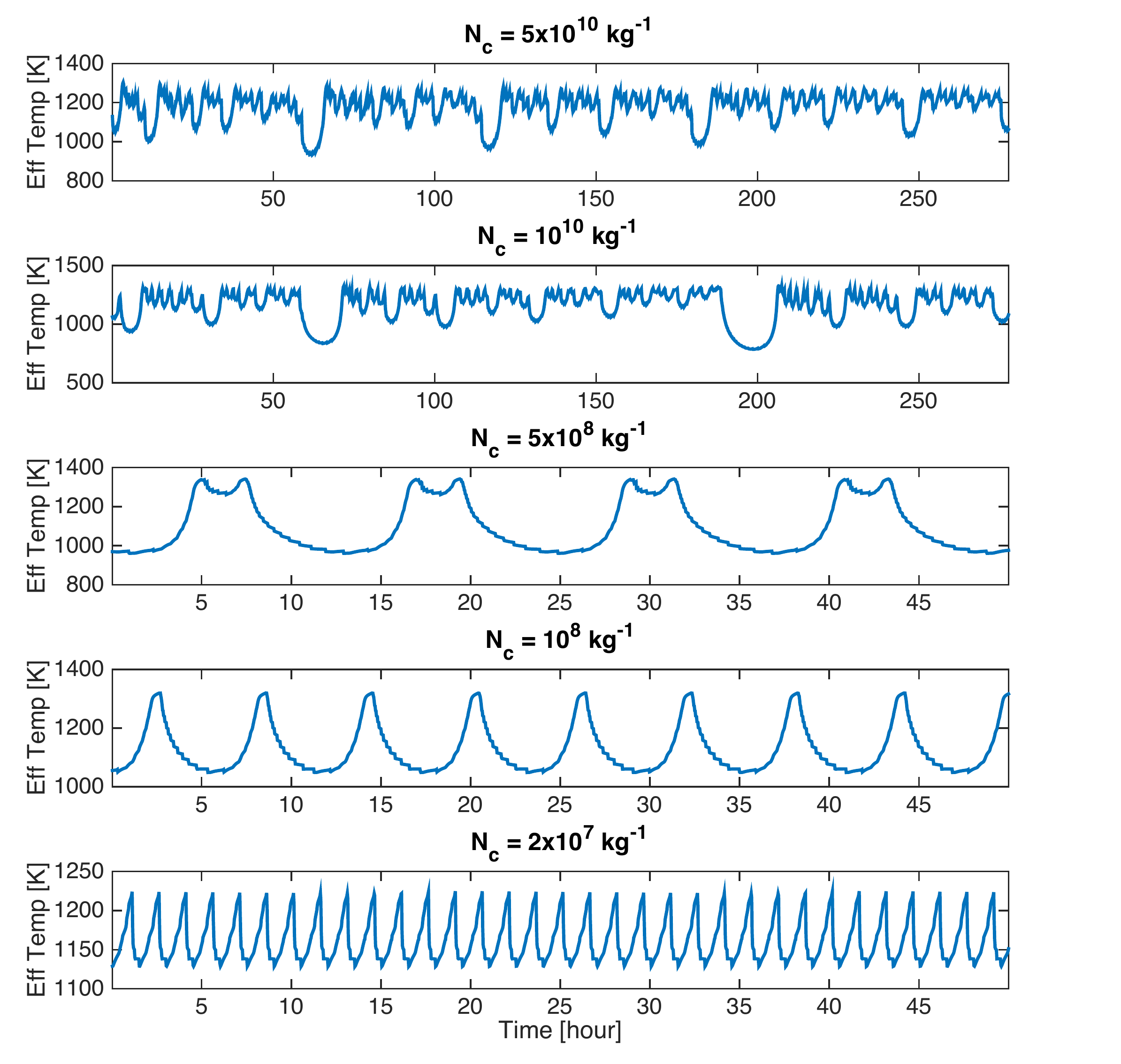}
\caption{Outgoing flux plotted as effective temperature as a function  of time for cases with cloud number density $\mathcal{N}_c=5\times10^{10}, ~ 10^{10}, ~5\times10^8$ \rm{(the nominal model)}, $10^8$ and $2\times10^7  ~ \rm{kg^{-1}}$. Other model parameters are the same as the nominal case in Section \ref{nominal}. 
}
\label{nseed}
\end{figure}

\subsection{Sensitivity Studies}
We study sensitivity of the variability to various model assumptions, and show that the detailed evolution is sensitive to model parameters, some of which in reality are rather unconstrained. However, the existence of variability is robust over a wide range of model assumptions. In the following sections, only the mentioned parameters are changed, and other parameters are the same as the nominal case. We discuss some important cases in details but leave  out details of most cases to avoid redundancy.

\subsubsection{Cloud Number Density}
\label{ch.Nc}
The particle number per mass  $\mathcal{N}_c$ controls the particle sizes and cloud opacities, determining the cloud settling flux and heating/cooling rate.  We perform a series of models with  $\mathcal{N}_c=2\times10^7$ to $5\times10^{10}  ~ \rm{kg^{-1}}$,  which is a significant variation around the nominal value $5\times10^{8} ~ \rm{kg^{-1}}$. To give a sense, assuming a cloud mixing ratio $10^{-3}$ kg/kg, the peak of the particle size distribution is at about 1.3 $\mu$m and 0.1 $\mu$m for $\mathcal{N}_c=2\times10^7$ and $5\times10^{10}  ~ \rm{kg^{-1}}$, respectively.  Figure \ref{nseed} shows the outgoing thermal flux in terms of effective temperature as a function of time. Strikingly, the high-$\mathcal{N}_c$  cases ($\mathcal{N}_c=5\times10^{10}$ and $10^{10}~ \rm{kg^{-1}}$) exhibit irregular variabilities as opposed to models with number density smaller than $5\times10^{8}~ \rm{kg^{-1}}$ that exhibit quasi-periodic variabilities. A clear trend for the low-number-density models is the decreasing oscillation period  with decreasing $\mathcal{N}_c$. The decrease of $\mathcal{N}_c$ causes the peak of size spectrum shift to larger particles and thus a larger settling flux.  If the change of size spectrum is gradual (i.e., the corresponding cloud extinction coefficient distribution does not change drastically), the decrease of $\mathcal{N}_c$ reduces the cloud opacity. Qualitatively, both effects reduce the time needed to complete a cloud cycle.

The  transition from regular to irregular variability with increasing cloud number density is due to the onset of chaos. The evolution of the system is sensitive to initial conditions when $\mathcal{N}_c$ is high.  We perform experiments  of slightly different initial conditions for the model with $\mathcal{N}_c=5\times10^{10}~ \rm{kg^{-1}}$ by  perturbing   just 1\% the cloud mixing ratio in one grid point. The initial condition of this experiment is taken from an instantaneous output of a  $\mathcal{N}_c=5\times10^{10}~ \rm{kg^{-1}}$ model after reaching a statistically equilibrium. The evolution of the two cases shows drastically different trajectories after some  time as shown in the upper right panel of Figure \ref{combine} in  Appendix \ref{ch.chaos}.    On the contrary, experiments with vastly different initial cloud structures for the model with $\mathcal{N}_c=10^{8}~ \rm{kg^{-1}}$ show a quick merging of the evolution to the regular periodic state after some differences at the very beginning.  The sensitive dependence of high- $\mathcal{N}_c$ models is the root cause of irregularity shown in the evolution of the systems (e.g., \citealp{motter2013}). 

\begin{figure*}      
\epsscale{0.9}      
\plotone{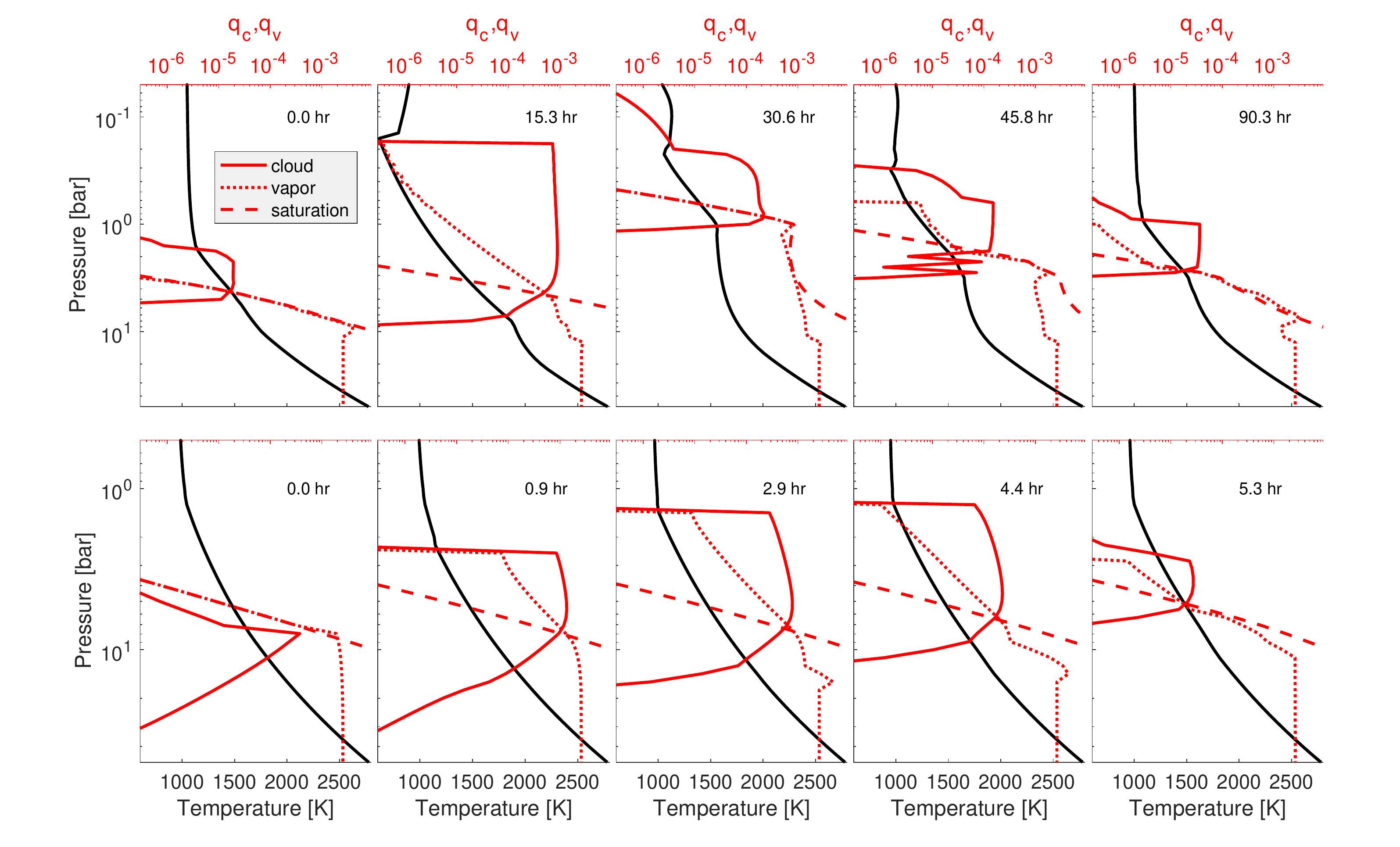}
\caption{Snapshots of temperature, vapor, cloud mixing ratio and saturation vapor mixing ratio for models with cloud number per mass   $\mathcal{N}_c=5\times10^{10}$ in the first row and $10^8 ~ \rm{kg^{-1}}$ in the second row  (a factor of 100 times greater and 5 times less than the nominal value, respectively).  The black curve represents temperature (bottom abscissa) and red curves represent cloud, vapor, and saturation vapor mixing ratios (top abscissa). Time increases from left to right starting from zero as shown in the panels.
}
\label{timeplot2}
\end{figure*}

Chaos can emerge from a forced-dissipated nonlinear system \citep{lorenz1963}. In our system, the qualitative condition under which chaos may occur is likely that the cloud radiative heating should be able to  strongly alter the temperature structure, such that the  \emph{saturation} vapor mixing ratio profile can evolve significantly.  On the other hand, in conditions showing quasi-regular variabilities, the variation of temperature near the cloud base is quite small, and so the  saturation vapor mixing ratio profile stays roughly unchanged.  In the latter condition,  the cloud cycle is controlled by the processes of settling and stratification (which only needs a slight change of thermal structure) which are well described by the ideal picture in the mechanism section \ref{ch.mechanism}. In essence, a new cloud cycle starts only when the stratification disappears.  In this sense, the system is simple and ``linear". The system can become sufficient ``nonlinear" when the saturation vapor mixing ratio can significantly evolve. Formation of clouds can then be triggered additionally by cooling of the thermal structure that supersaturates local vapor. The varying saturation vapor mixing ratio serves as an extra degree of freedom to the system which is nonlinearly coupled to other variables, leading to strong ``nonlinearity", a preferred condition for the emergence of chaos.   To illustrate the above two conditions,  Figure \ref{timeplot2} presents several snapshots of the temperature, cloud, vapor and saturation vapor profiles for the  model with $\mathcal{N}_c=5\times10^{10}~ \rm{kg^{-1}}$  in the upper row and for the model with $\mathcal{N}_c= 10^{8}~ \rm{kg^{-1}}$ in the lower row. Because the model with $\mathcal{N}_c=5\times10^{10}~ \rm{kg^{-1}}$ is irregular, the snapshots are randomly chosen; but the snapshots for the model with $\mathcal{N}_c=10^{8}~ \rm{kg^{-1}}$ sample a full single cloud cycle.  We can see obviously that in the large-$\mathcal{N}_c$ model, all profiles evolve substantially. Below the cloud base, the vapor profile is sometimes quite close to the saturation vapor profile, so that any slight cooling can trigger cloud formation (for example, the spiky cloud structure seen below the main cloud deck in the fourth upper panel).    On the contrary, in the low-$\mathcal{N}_c$ model, the saturation vapor pressure curve barely varies. In fact,  all models explored in this study are either quasi-regular or irregular variability, and their evolution all fall into either the two categories shown in Figure \ref{timeplot2}.

\begin{figure}      
\epsscale{1.25}      
\plotone{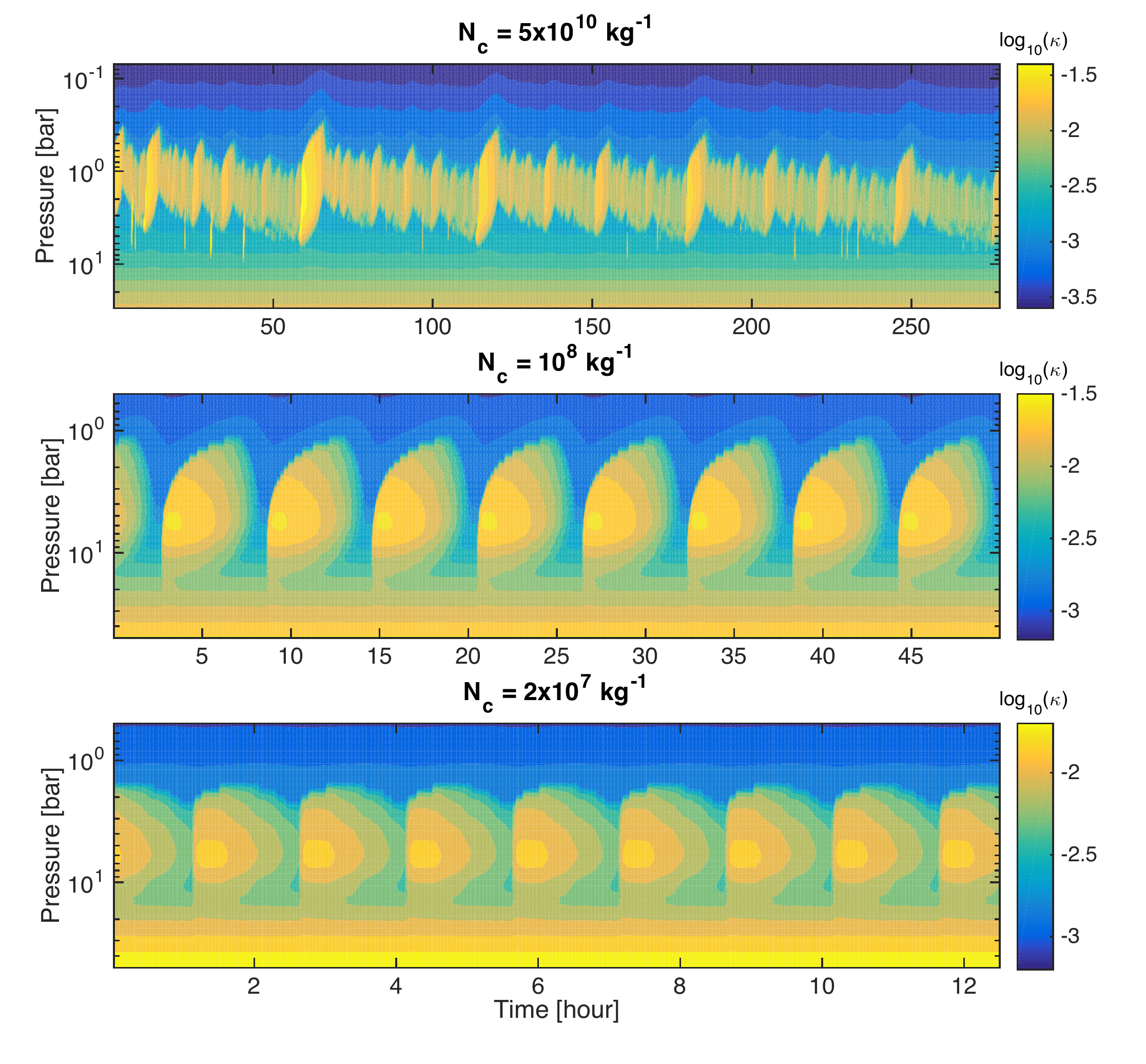}
\caption{Time evolution of the total opacity as a function of pressure in logarithmic scale for models with cloud number density $\mathcal{N}_c=5\times10^{10}$ (upper panel), $10^8 ~ \rm{kg^{-1}}$ (middle panel) and $2\times10^7 ~ \rm{kg^{-1}}$ (lower panel). 
}
\label{cloudcompare}
\end{figure} 

Interestingly,  high-$\mathcal{N}_c$ models show cloud coverage almost all the time during the evolution as apposed to the relatively-low-$\mathcal{N}_c$ models in which cloud opacities sometimes contribute negligibly to the total opacity. We show the total opacity structure as a function of time in Figure \ref{cloudcompare} for the models with $\mathcal{N}_c=5\times10^{10}$ in the upper panel, $10^8 ~ \rm{kg^{-1}}$ in the middle panel and $2\times10^7 ~ \rm{kg^{-1}}$ in the lower panel. The background gaseous opacity generally decreases with decreasing pressure, and so it is easy to recognize the cloud opacity as those local   maximums evolving near the cloud forming region. The high-$\mathcal{N}_c$ model show no cloud opacity gap in the evolution, while there is obvious opacity gap   in the medium-$\mathcal{N}_c$ model, but then shows almost no gap again in the low-$\mathcal{N}_c$ model. Qualitatively, cloud cycles in the high-$\mathcal{N}_c$ models do not rely on a complete dissipation of clouds but  instead are mainly driven by heating/cooling of the thermal profile, so that we rarely see a cloud-free instant in these models. In the medium-$\mathcal{N}_c$ model, cloud opacity has a much smaller impact on the thermal structure, and thus a nearly full dissipation of cloud deck is allowed. In the case of very low  $\mathcal{N}_c$, duration of stratification is too short because of less cloud radiative warming, and the new cloud cycle is quickly triggered before the old cloud deck fully dissipates. We will discuss the implications in the discussion section.

\subsubsection{The Background Tracer Diffusion}
\begin{figure}      
\epsscale{1.25}      
\plotone{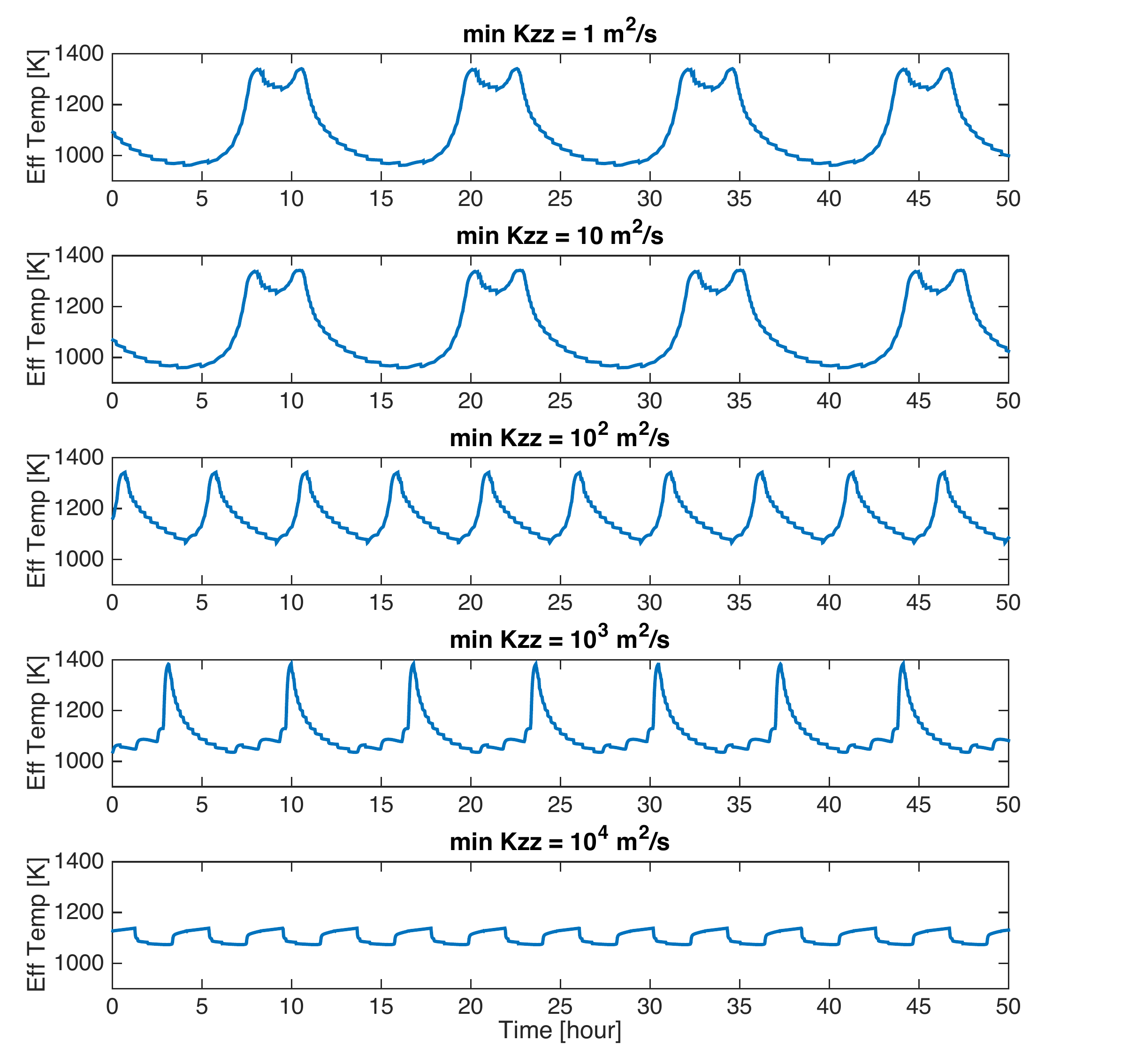}
\caption{Outgoing thermal flux in terms of effective temperature as a function of time for models with a constant minimal  tracer diffusion coefficient $K_{\rm{zz, min}} = 1 , 10 , 10^2, 10^3$ and $10^4~\rm{m^2s^{-1}}$. Other parameters are the same as the nominal model.
}
\label{kzzflux}
\end{figure}

Vertical transport of tracers could occur in the stratified atmospheres via, for instance, shear instability, wave breaking and large-scale winds. It is difficult to quantify these effects without detailed understanding of the atmospheres. Instead  we explore the effect of mixing in the stratified layer   by  simply imposing an additional constant background diffusion $K_{\rm{zz, min}}$ for tracers, such that the total diffusion coefficient is $\max [K_{\rm{zz}}, K_{\rm{zz, min}}]$ with the former determined by Eq. (\ref{eq.kzz}). We explore values  from $K_{\rm{zz, min}} = 1$ to $10^4~\rm{m^2s^{-1}}$. Figure \ref{kzzflux} shows their effective temperature as a function of time. All models show variability in the thermal flux. Cases with relatively small  $K_{\rm{zz, min}}$ (1 and $10 ~\rm{m^2s^{-1}}$) have both  oscillation periods and flux variation amplitudes quantitatively similar  to the nominal case. Cases with medium values ($10^2$ and $10^3~\rm{m^2s^{-1}}$)  display quantitatively different evolutions but still retain large variation and somewhat close oscillation periods as the nominal model. The case with  $K_{\rm{zz, min}}=10^4~\rm{m^2s^{-1}}$ exhibits a higher oscillation frequency and a much smaller variation amplitude. 

We conclude that a medium background diffusion  is unlikely to suppress the variability under parameter regime of the nominal model. However, the details can  be affected. Roughly speaking,  nontrivial background diffusion can efficiently mix deep vapor upward and constantly promote cloud formation near the cloud base. This is different from the nominal model where supply of deep vapor is terminated once the atmosphere below the cloud base is stratified. Compared to the nominal model, this brings the temperature profile below the cloud base closer to the condensation curve and uplifts the cloud base to a higher altitudes because of cloud radiative heating. Particle settling strength is higher at lower pressure, and thus partly contributes to the subtly different evolutionary details seen in Figure \ref{kzzflux}. For sufficiently large background diffusion that can completely balance the particle settling, the variability could be suppressed and the system can approach a steady state.     Values of the diffusion coefficient $K_{\rm{zz}}$ derived from fitting to spectrums of field BDs using non-equilibrium chemical models are moderate  ($ \sim 10^2 ~\rm{m^2s^{-1}}$ or $ \sim 10^6~\rm{cm^2s^{-1}}$, e.g., \citealp{stephens2009}) for many L and T dwarfs, which is unlikely to suppress variability in the parameter regime similar to our nominal model.

\begin{figure}      
\epsscale{1.25}      
\plotone{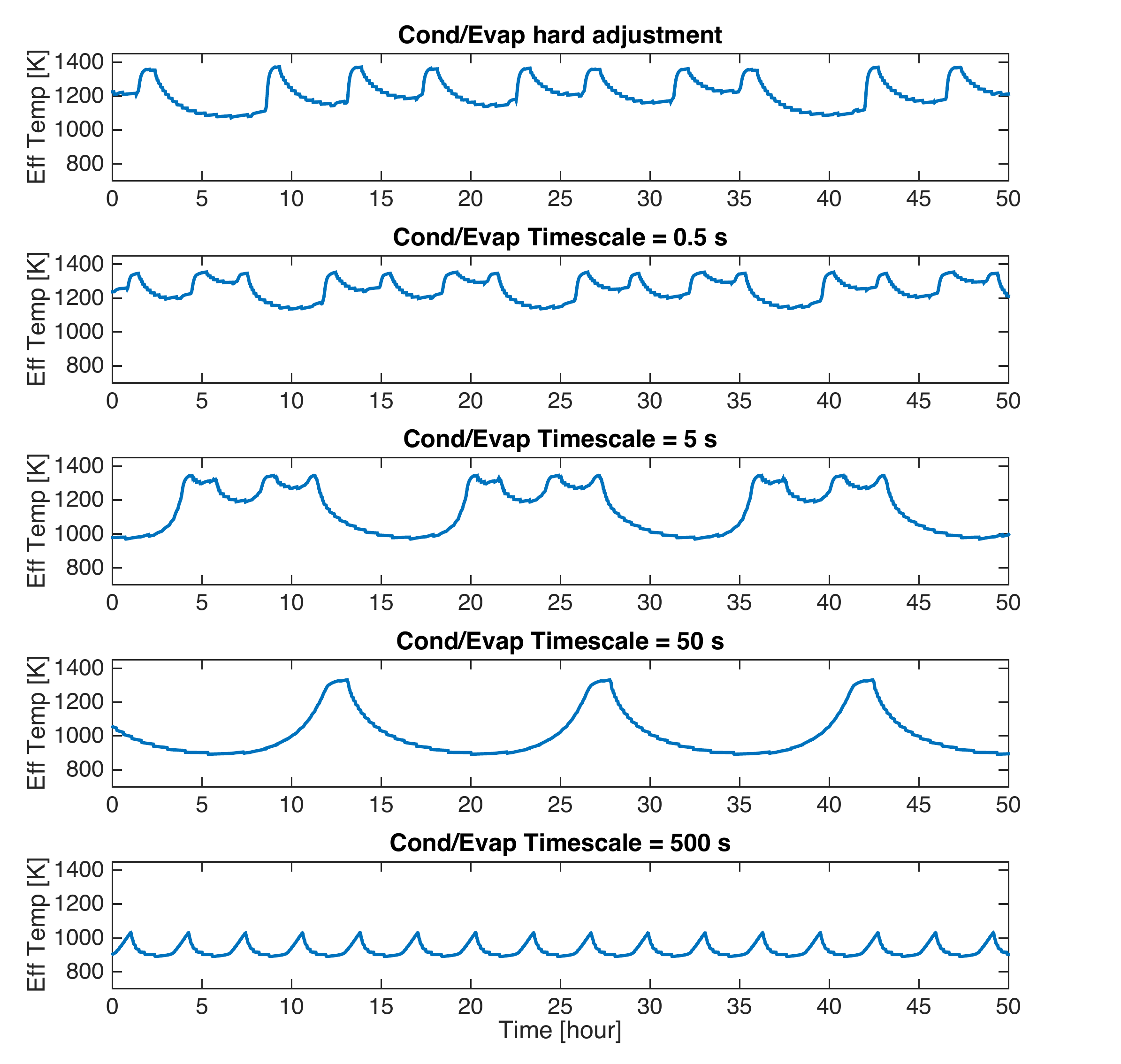}
\caption{Outgoing thermal flux in terms of effective temperature as a function of time for cases with different conversion timescales $\tau_c=\rm{infinitely~small}, 0.5, 5, 50$ and 500 s. In the case of $\tau_c=\rm{infinitely~small}$, we did hard adjustment for conversion, i.e., vapor/cloud mixing ratio is instantaneously adjusted such that no supersaturation/clouds-in-unsaturated-air is allowed.
}
\label{taucond}
\end{figure}

\subsubsection{Conversion Timescale}
One interesting feature in the nominal model is that the supersaturation can be orders of magnitude larger than 100\% (see Figure \ref{mechanism}).  In earth's atmosphere, the supersaturation of water seldom exceeds 1\% (e.g., \citealp{houze2014}) because proper cloud condensation nucleus (CCN) enhance the nucleation process. Even without proper CCNs, the homogeneous nucleation rate increases nonlinearly with increasing supersaturation (e.g., chapter 3, \citealp{fletcher2011}), and tends to prevent the atmosphere arriving at such high supersaturation\footnote{For instance, $\sim$300\% supersaturation is sufficient to trigger efficient homogeneous nucleation of water vapor in earth's troposphere (e.g., \citealp{houze2014}).}. However, this may not be the case in brown dwarfs' atmospheres \citep{helling2013}, and detailed cloud microphysics models suggest  that supersaturation up to several orders of magnitude larger than unity is possible in these atmospheres (see Figure 9 in \citealp{helling2014}). In any case, we examine whether the existence of  variability relies on high supersaturation, and how  variability behaves with varying supersaturation. This is done simply by adjusting  the conversion timescale $\tau_c$ which mainly controls the degree of supersaturation as shown in Eq. (\ref{eq.supersaturation}). Figure \ref{taucond}   shows outgoing flux from models with $\tau_c=\rm{infinitely~small}, 0.5, 5, 50$ and 500 s.  In the case of infinitely small $\tau_c$, we perform a ``hard'' adjustment for conversion between vapor and cloud. This means that whenever there is super saturation for vapor, or cloud in a subsaturated environment, we instantaneously adjust the vapor and cloud to an equilibrium state without using a relaxation scheme.  Variability emerges  from all models, suggesting that the degree of supersaturation within a  reasonable range does not affect the existence of variability.  Models with $\tau_c \leq 50$ s, although differing in quantitative details, exhibit qualitatively similar temporal evolution. This is  because the strong convective mixing is able to mix the bulk  tracers ($q_v+q_c$) against cloud settling, and the detailed conversion between vapor and clouds does not affect the overall mechanisms driving the variability.  The model with $\tau_c = 500$ s shows smaller average flux and variation amplitude. In this case, abundant vapor can be maintained throughout the cloud-forming region almost all the time. Thus, the main cloud deck is thick and has little variability due to supply from the invariant vapor profile. The main variability contributing to the outgoing flux is by the slightly varying cloud-top altitudes.

\begin{figure}      
\epsscale{1.1}      
\plotone{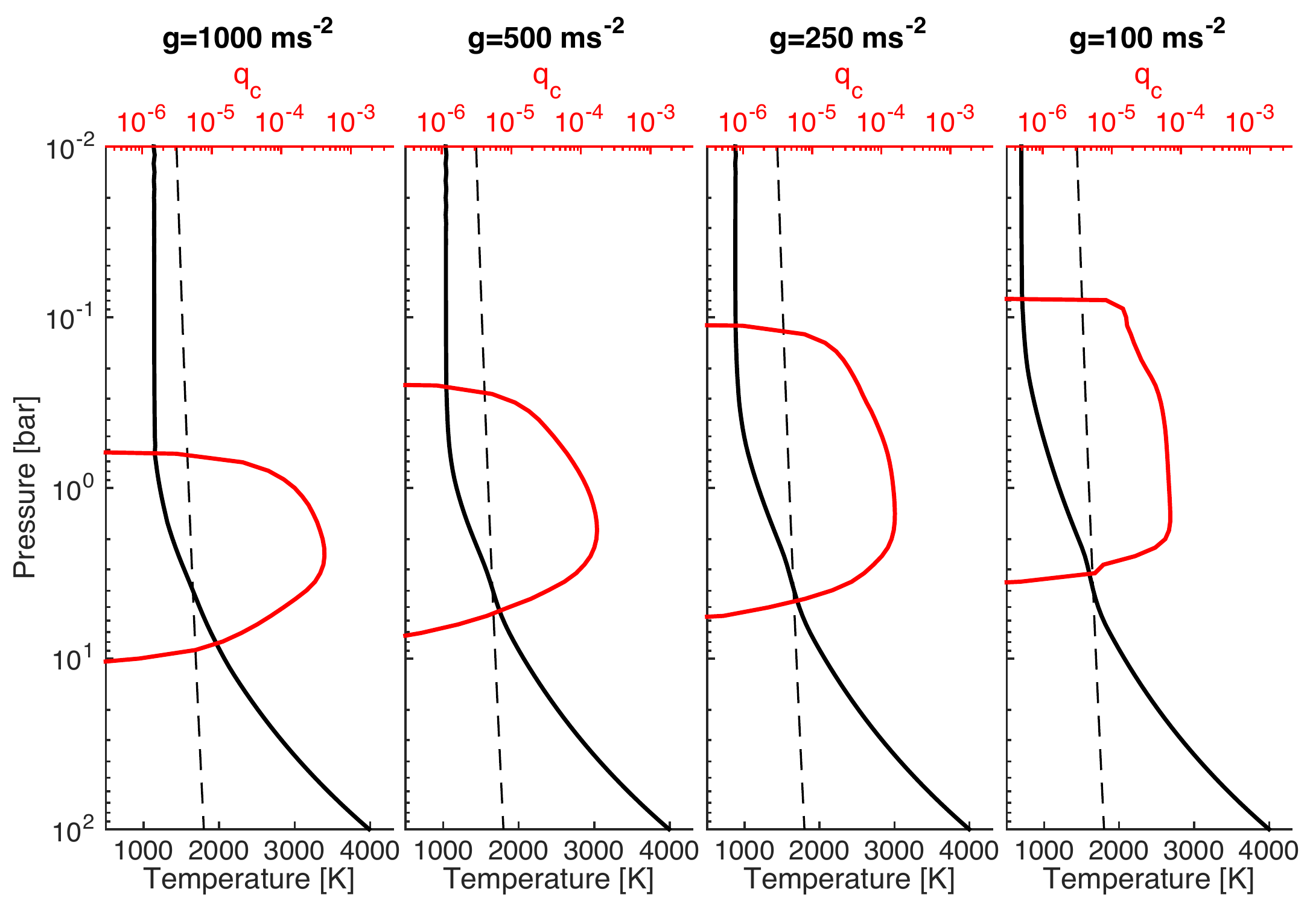}
\caption{Time-averaged temperature and cloud structure for models with different surface gravity $g=1000, 500, 250$ and $100 ~\rm{m~ s^{-2}}$, and all models have a temperature 4000 K at the model bottom boundary 100 bars. 
}
\label{gravitytp}
\end{figure}

\begin{figure}      
\epsscale{1.25}      
\plotone{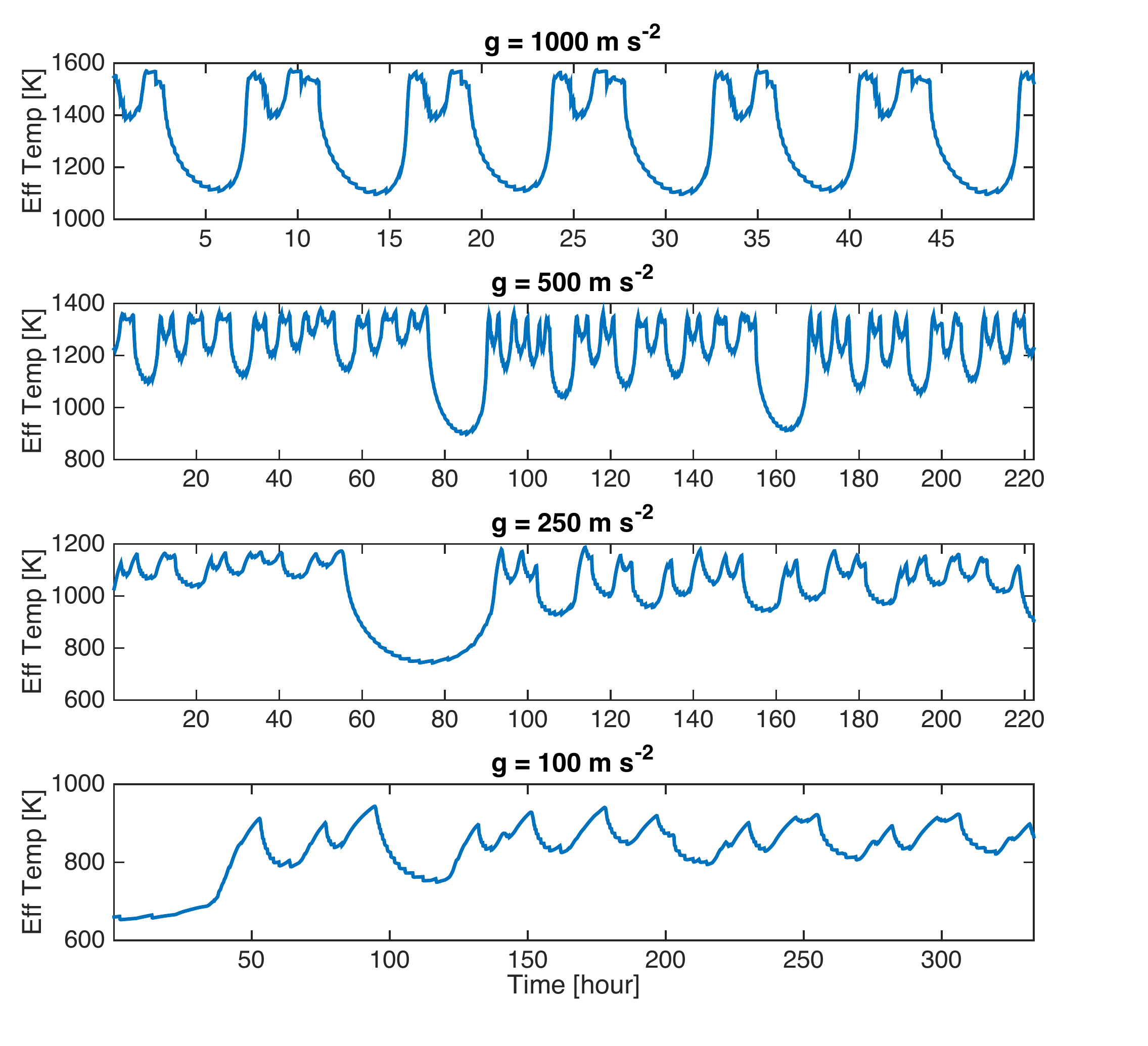}
\caption{Outgoing thermal flux in terms of effective temperature as a function of time for models with different surface gravity $g=1000, 500, 250$ and $100 ~\rm{m~ s^{-2}}$ of models have a temperature of 4000 K at the model bottom boundary of 100 bars. 
}
\label{gravityflux}
\end{figure}

\subsubsection{Surface Gravity}
Variability has also been detected in relatively low-gravity objects, including directly imaged  EGPs and free-floating  planetary-mass objects \citep{zhou2016, biller2015, gagne2017}, although it is yet unclear for these objects whether the surface features evolve on a short timescale in a similar fashion as many field brown dwarfs \citep{metchev2015, apai2017}.  Changes in the surface gravity  affect the cloud settling velocity, convective mixing  and the T-P structure by changing the optical depth. But we expect that the changes of gravity does not affect the fundamental driving mechanisms of variability.   We carry out simulations with gravity $g=1000, 500, 250$ and $100 ~\rm{m~ s^{-2}}$, and all models have a temperature 4000 K at the model bottom boundary 100 bars.  Figure \ref{gravitytp} shows the time-average T-P and cloud mixing ratio profiles and Figure \ref{gravityflux} shows the outgoing flux  for the models. If there there were no radiative cloud feedback, the pressure of the cloud base would be the same for all models because the intersections of the T-P profile and the condensation curve are in the convective layer. With decreasing gravity, the atmospheres undergo  greater back  warming below the cloud base due to the lower settling velocity in lower-gravity conditions. This results in the higher-altitude time-average cloud base in the lower-gravity cases. As discussed in Section \ref{ch.mechanism}, the altitude of cloud top is roughly where the optical depth transition to $\ll 1$. If the background gaseous opacity is not very sensitive to temperature, the pressure of the cloud top roughly scales as gravity, which can loosely explain the trend of time-average cloud top pressures (0.6, 0.27, 0.13, 0.08 bar for $g=1000, 500, 250$ and $100 ~\rm{m~ s^{-2}}$, respectively) shown in Figure \ref{gravitytp}. The out-going flux of the $g=1000~\rm{m~ s^{-2}}$ model is quasi-periodic, but the fluxes are irregular for lower-gravity models due to the onset of chaos. As discussed above, the onset of chaos in low-gravity models is due to the  sufficient temperature change by cloud radiative heating, as demonstrated  in Figure \ref{gravitytp}.

\subsubsection{Atmospheric Temperature}
\begin{figure}      
\epsscale{1.25}      
\plotone{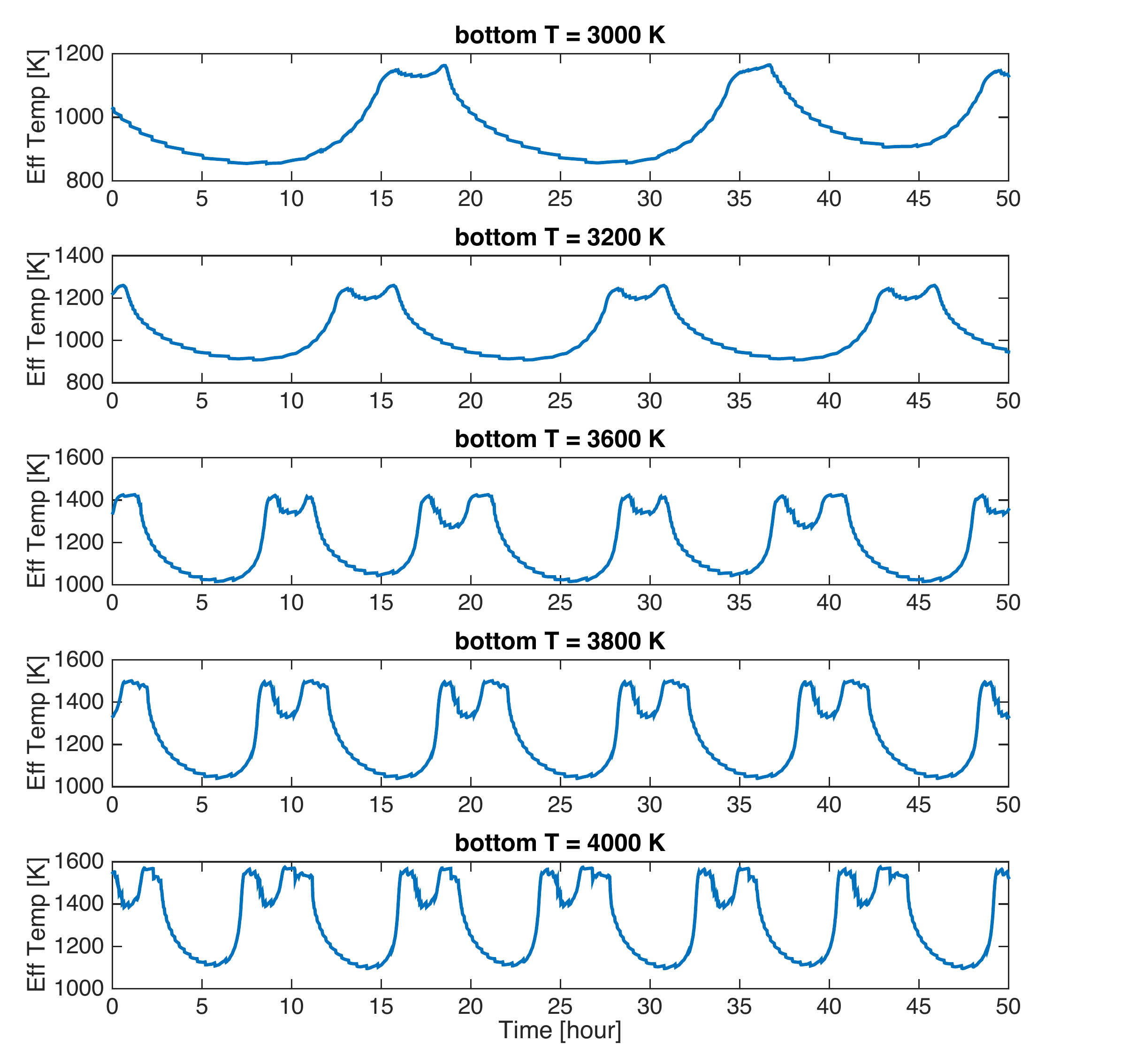}
\caption{Outgoing thermal flux in terms of effective temperature as a function of time for models with different temperatures $T_b = 3000, 3200, 3600, 3800$ and 4000 K  at the bottom boundary 100 bars.
}
\label{tbot}
\end{figure}

Variability is common among field L and T dwarfs across a wide range of atmospheric temperatures. Different temperature affects the pressure level of first condensation, and thus the cloud settling velocity and radiative heating/cooling rate. Here we show experiments in which  we change the temperature $T_b = 3000, 3200, 3600, 3800$ and 4000 K (note that 3400 K is the value adopted for the nominal model)  at the bottom boundary 100 bars in Figure \ref{tbot}. All the models show quasi-periodic variability and an obvious trend: the oscillation frequency is higher for hotter models. Qualitatively, two factors are responsible. First, in hotter models, the cloud layer dissipates faster after it forms due to the lower-pressure cloud base and thus the higher settling velocities. Second, in hotter models, the cloud-forming region cools down faster after cloud dissipates due to the shorter thermal relaxation timescale, helping to restore the next cloud cycle in a shorter timescale.

\subsubsection{Shape of Size Distribution Function}
\begin{figure}      
\epsscale{1.25}      
\plotone{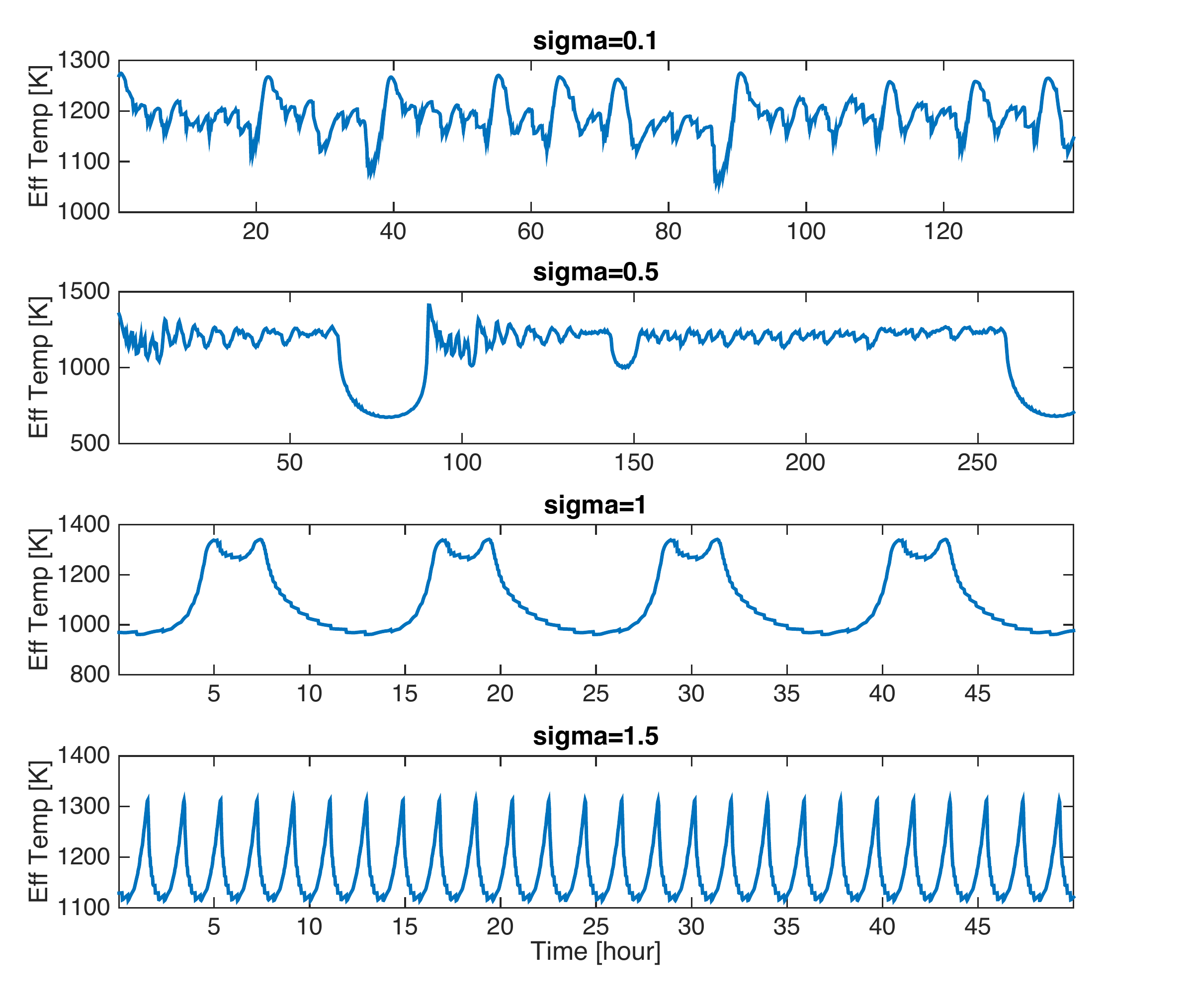}
\caption{Outgoing thermal flux in terms of effective temperature as a function of time for models with $\sigma = 0.1, 0.5, 1$ and 1.5. Here $\sigma$ is a parameter controls the width of cloud particle size distribution.
}
\label{sigmalog}
\end{figure}

We perform a sequence of models varying parameter $\sigma$ controlling the width of the log-normal size distribution of cloud particles  to examine the sensitivity to the shape of size distribution function. Figure \ref{sigmalog} shows outgoing fluxes of models with $\sigma = 0.1, 0.5, 1$ and 1.5. Models with relatively narrow  size distribution ($\sigma=0.1$ and 0.5) show irregular variability. In these cases, the settling flux is relatively small compared to large-$\sigma$ models because of the smaller fraction of large particles. The model with $\sigma=1.5$ is regular and oscillates with a much shorter period than the model with $\sigma=1$ due to a much larger settling flux. As discussed in Section \ref{model}, models with exponential cloud size distribution also exhibit variabilities qualitatively similar to the log-normal distribution. We conclude that details of the cloud size distribution strongly affect  the quantitative behavior  of the variability, but the mechanisms remain the same. 

\section{Discussion}
\label{discussion}

\subsection{Under What Conditions Does the 1D Cloud-driven Variability   not Occur?}
It is impractical to explore the variability through the whole parameter space, and we qualitatively discuss   conditions under which the 1D system governed by Eqs. (\ref{eq.temp}) to (\ref{eq.qc}) may have little variability.  We list several, but probably not all, such conditions. \emph{1, Vapor first condense in the stratified layer, no background mixing and relatively large cloud particles}: Under these conditions, the cloud layer will quickly dissipate via particle settling, but cannot be replenished by vapor transport because the atmosphere is stratified and there is no background mixing. \emph{2, Strong background  mixing and small cloud particles}: very strong background mixing is able to completely balance particle settling, suppressing the cloud dissipation cycle. \emph{3, Condensable vapor mixing ratio is too small}: in this case clouds may be too optically thin to effectively alter the thermal structure and maintain the cloud cycle. Considering realistic atmospheric conditions from mid L to mid  T dwarfs including directly imaged EGPs, the conditions leading to no variability seem stringent. Perhaps late T dwarfs represent conditions under which cloud-driven variability does not easily occur, at least in our simple 1D context, because  major condensates such as silicates and iron condense at deep layers where the background gaseous opacity is too large for cloud opacities to matter much. Indeed, cloud-free models can interpret spectra of late T dwarfs reasonably well (e.g., \citealp{line2017}). This suggests that our mechanism should apply to the majority of brown dwarfs and directly imaged EGPs for which reasonably thick clouds are expected to exist in the observable atmospheres.

  \subsection{Implications for Observations}
The radiative-cloud-driven variability provides a  theoretical foundation to understand the rapidly evolving light curves found for many L and T dwarfs (e.g., \citealp{artigau2009, metchev2015, apai2017}). These observations are difficult to explain by rotational modulation caused by temporally steady or slowly evolving surface inhomogeneity  \citep{karalidi2016}, for example, phenomena analogous to the Great Red Spot or 5-$\mu$m hot spots in Jupiter's atmosphere.  Light curves of many brown dwarfs can only be explained when the surface features evolve over timescales comparable to the rotation period. Our results show that the timescale of the cloud-driven variability can be as short as a few hours depending on model parameters. If the surface features are driven by such variability,  the rotation-modulated light curves will be irregular over rotational timescales. Moreover, even if the BD is observed from a nearly pole-on geometry---which minimizes the effect of rotational modulation in the light curves---short-term evolution of light curves could still occur due to the evolving statistics of surface patches driven by variability caused with the mechanism explored here. Nearly pole-on variable BDs could be valuable targets to characterize the intrinsic variability driven by clouds. 
 Our models  show that thick clouds usually correlate to a convective and cooler profile, and thin clouds correlate to a stratified and warmer profile, which is a result of coevolution of clouds and thermal structure.  For some  variable BDs   characterized at multiple wavelengths, a combination of two types of surfaces consisting a  warm, thin-cloud patch and a cool, thick-cloud patch is necessary to fit the observations \citep{apai2013, buenzli2015b}. Our model provides a mechanism for such a combination of surface features.
In our model, temperature variation exhibits a pressure-dependent shift and the maximum difference is between the cloud top and base (see the upper panel in Figure \ref{timeplot1}). This may shed light on the observed offsets of wavelength-dependent light curves for some variable BDs \citep{buenzli2012,  yang2016}.  Our simple grey model here only reveals the mechanisms, and more sophisticated non-grey models with proper chemistry are needed to quantify the spectroscopically resolved variability.

The characteristic evolution of clouds in models with relatively large particle sizes (see Figure \ref{timeplot1} and \ref{cloudcompare}) provides a mechanism for cloud breaking,  in which clouds  sometimes almost fully dissipate and contribute negligibly to the total opacity. Statistically, the surface cloud patches governed by such cloud cycles must have some portion occupying the cloud-dissipated state. As a result,  the globally integrated outgoing flux of the atmosphere is only partially affected by cloud opacity. On the other hand, in models with relatively small particle sizes, full cloud dissipation does not easily occur, and the cloud opacity always contributes significantly to the total opacity. As a result, the integrated outgoing flux of atmospheres with small particles are likely significantly affected by cloud opacity. This has important application to the longstanding  issue associated with the L to T dwarf transition that shows a sudden change of near-IR colors over a narrow effective temperature range,  the \emph{J}-band brightening and resurgence of the gaseous  FeH bands  (e.g., \citealp{kirkpatrick2005, burgasser2002}). Cloud breaking remains the most promising mechanism to explain the L/T transition  \citep{ackerman2001, burgasser2002, marley2010}.  However, exactly why and how clouds break remains unclear.  The working mechanisms and detailed evolution of cloud cycles shown in this work support this idea   in a natural way:  by interpreting observations using atmospheric and radiative transfer models, it becomes clear that  early to mid L  dwarfs are likely dominated by sub-micron particles (e.g., \citealp{burningham2017, hiranaka2016, allard2001}), and T dwarfs are reasonably well represented by large-cloud-particle models (e.g., \citealp{saumon2008}). Given that the  transition from L to T spectral type is accompanied with change from small to large particles, cloud breaking starting at the L/T transition is then a natural outcome from this sequence, as cloud breaking in time variability operates only when particles becomes relatively large. The condition for cloud breaking  depends primarily on particles size but is rather insensitive to atmospheric temperature. Thus, in principle, the onset of cloud breaking can occur over a narrow effective temperature range.   In addition, our models with larger particles usually have cooler atmospheres above the condensation level than those with smaller particles, promoting the conversion from CO to $\rm{CH_4}$. 
   
   \subsection{Comparison with Other Cloud Models}
   \label{ch.othercloudmodel}
  The most distinctive features of our 1D model compared to other prevalent parameterized cloud models for substellar atmospheres \citep{ackerman2001, allard2001, tsuji2002, cooper2003, burrows2006, charnay2018} are the relaxation of exact balance on both thermal structure and clouds at any given time, and self-consistent coupling between thermal structure, mixing and clouds. These treatments bring new insights to understanding clouds in substellar atmospheres.  Our results demonstrate that properties of a single cloud layer can have substantial variation over a short timescale, with major changes in the cloud mass loading, altitudes of the cloud base and top and thus the layer thickness, and  the particle size distribution.   The two-dimensional hydrodynamic  cloud formation model by \cite{freytag2010} is a good point of reference; unfortunately,  \cite{freytag2010}  explored only regimes of small particles, and did not discuss details of the variability, preventing  comparisons.  As discussed in Section \ref{nominal},  a single set of  statistically-averaged  temperature and cloud profiles  cannot  represent the equilibrium state of atmospheres with vigorous cloud formation.  Retrieval  methods (e.g., \citealp{line2017}) have shown that a single set of radiative-convective equilibrium profiles can well represent spectra of many late T dwarfs whose atmospheres are relatively cloud-free. It would be interesting to extend this type of study to samples of late L and early T dwarfs.

A  first-principle cloud microphysics approach  by Helling et al. in a series of papers (here we refer to Helling's model, e.g., \citealp{helling2001, helling2008, lee2016, lines2018}, see \citealp{helling2013, helling2014} for   reviews) considers that CCN are available only through homogeneous nucleation of certain species ($\rm{TiO_2}$ in their models), and require extremely high supersaturation, i.e., the nucleation occurs at much higher altitudes above the condensation level predicted by the equilibrium chemistry. Clouds then form through heterogeneous condensation of various condensable gases upon $\rm{TiO_2}$ seed particles that fall down from the upper atmosphere. The resulting cloud structure consists of ``dirty grains'' with a continuous distribution from upper down to deeper atmosphere until the temperatures become sufficient for the particles to evaporate. A crucial assumption in their model is the continuous, vigorous convective transport  of  condensable vapor upward into nucleation regions regardless the atmospheric conditions---regardless of whether it is stratified or convective \citep{woitke2004}.  As shown in this study, an important consequence of cloud formation is the stratification below the cloud base and the suppression of upward convective mixing of vapors (though some mixing may still be possible through eddies in the stratified zone). This should have a significant impact on the nucleation rate in the upper atmosphere and thus the consequent cloud formation following the seed particles. One can imagine that  the nucleation in the upper atmosphere can then evolve following the evolving stratification below the cloud base, which should add additional complexity to the variability, beyond the mechanisms explored in this study. Along a similar line, \cite{gao2018} and \cite{powell2018} presented  bin-resolved cloud microphysics models, showing that convective mixing is an important parameter controlling the cloud properties. Likewise, our results suggest that the variability should have a significant impact on the detailed cloud microphysics in brown dwarf atmospheres. Interestingly, using microphysical models, \cite{gao2014} showed that cloud properties in Venus' atmosphere could exhibit long-term variability (over timescales of months), and the variability is caused by mechanisms in the microphysical level. If such microphysical mechanisms occur in BDs and directly imaged EGPs, the coupling to the mechanism presented in this study would suggest a much more complicated situation in these atmospheres.

In reality clouds in brown dwarfs may be composed of several cloud layers with different compositions  as argued by equilibrium chemical  models (e.g., \citealp{visscher2006, visscher2010}).  We expect that this would not suppress the variability, which we have explored in this work using only the condensation properties of the enstatite ($\rm{MgSiO_3}$) cloud. From mid-L to mid-T dwarfs, the most abundant cloud species that dominate the heating/cooling   are silicate and iron clouds (e.g., \citealp{ackerman2001}), whose condensate pressures are not far away from each other. Once a detached convective zone is formed by radiative cloud heating, the silicate and iron cloud layers are expected to evolve together. 

 Possibly more ``blurred'' cloud structures could occur, with particles comprising ``dirty grains'' comprising both silicates and iron, as suggested by detailed microphysical models by Helling's group as discussed above.  We argue that this chemical complexity would not suppress the variability. Even though the cloud structure in Helling's model could be smoother than the multi-layer cloud models predicted by equilibrium chemistry, a sharp transition to nearly cloud free will still occur near the base of the cloud layer (see figures in, e.g., \citealp{helling2008b}) presumably due to evaporation of silicate and iron components. This is the same as our model in which a sharp transition of cloud base can result in a detached secondary convective zone.  Meanwhile, the sharpness and location of the cloud top in our model are determined by interactions between convection and cloud radiative effect, but not by the composition of the clouds.  Therefore, we expect that the sharp cloud top, and thus the variability, would still exist even if the cloud microphysics is treated similarly as in Helling's model.

\subsection{Implications for Global Atmospheric Circulation}

One of the key interests on atmospheric circulation of brown dwarfs and directly imaged EGPs is whether large-scale flows are dominated by zonally banded structure similar to Jupiter and Saturn.  Classical two-dimensional turbulence theory predicts that banded structure can emerge from interactions between isotropic turbulence and the planetary rotation (see reviews by, e.g., \citealp{showman2018review, vasavada2005}).  In the context of cloud-free brown dwarfs, radiation is usually thought of as a form of thermal damping, in that hot (cold) regions tend to radiatiate greater (lesser) flux to space, lessening the amplitude of thermal perturbations relative to the mean radiative equilibrium state \citep{showman&kaspi2013, zhang&showman2014}.  If sufficiently strong, this radiative damping can damp out the atmospheric energy before it has time to reorganize into a zonally banded pattern comprising zonal jets \citep{zhang&showman2014}.
Our study suggests however that the existence of clouds can modify this picture.  In the presence of clouds, the local atmospheric temperature can undergo large variations over rapid timescales.  The variation is unlikely to be globally isotropic but is nevertheless likely to be patchy on regional-to-large scales. Such patches may each experience cloud-radiative feedbacks analogous to those captured in our 1D model, but with different phasing of their evolution throughout the cloud cycle.  This will contribute both to cloud patchiness that will help explain the observed IR lightcurve variability, and also result in temperature differences on isobars up to hundreds of Kelvins depending on cloud properties. As such, in addition to acting as as a strong damping, cloud effects cause radiation to act as a patchy, potentially random {\it forcing} of the atmosphere.
The size of a local patch that can be described by the 1D model is determined by large-scale dynamics in response to the multi-hundred K temperature perturbations. A natural dynamical length scale that might be relevant is the Rossby deformation radius over which atmospheric motions are strongly affected by planetary rotation.  Future 3D global models with cloud radiative feedback are necessary to investigate the large-scale dynamics that results from cloud feedback as well as the self-organization of the surface patches.

The strong influence of  radiative cloud feedback on the thermal structure raises an interesting question: how are the cloud effects coupled with the large-scale dynamics in BDs and directly imaged EGPs? In earth's tropics, deep moist convection driven by condensational latent heating and the associated cloud formation strongly control the local temperature structure. Large-scale equatorial disturbances are triggered by and coupled with the moist convection, manifesting themselves as zonally propagating waves covering a broad range of the spectrum in both zonal wavenumber and frequency  \citep{kiladis2009}. These  convectively coupled equatorial waves imprint rich and dominant signatures on the equatorial variability in the outgoing long-wave radiation \citep{wheeler1999}.  If the fundamental dynamical mechanisms driving the  coupled equatorial waves work  the same way in atmospheres of BDs and directly imaged EGPs, it is not surprising that similar waves triggered by cloud formation could exist in these atmospheres and be able to strongly affect  the observed variability.  Recently,  \cite{apai2017} showed that the puzzling evolution of long-term light curves of several BDs can be adequately explained by beating patterns caused by a pair of zonally propagating brightness disturbances that have different phase speeds, based on which they suggested the presence of banded structures similar to Neptune's atmosphere in these variable BDs. Here we imagine an alternative possible mechanism as the presence of coupled equatorial waves triggered by radiative cloud feedback. In this mechanism, the differential propagating brightness variation could be contributed by two dominant modes  of the waves. Banded structure is not a necessity because the differential propagating waves can be clumped at the equator to produce the same observational signals as those from banded structures. Global atmospheric models are necessary to demonstrate the viability of these ideas. Also, retrieval models are needed to probe the temperature and clouds structures associated with brightness anomalies to test if this mechanism is at play.

\section{Conclusions}
\label{conclusion}
Clouds  remain  one of the biggest obstacles in understanding ultra-cool atmospheres including those of brown dwarfs and directly imaged extrasolar giant planets.  In particular, clouds are thought to be the major cause of observed light curve variabilities of many BDs and directly imaged EGPs, as well as the longstanding, puzzling L to T dwarf transition.  In this study, we have investigated the short-time evolution of clouds and   thermal structures driven by radiative cloud feedback using a simple time-dependent 1D model that self-consistently treats cloud formation/dissipation, convective mixing and radiative transfer. We conclude that:
\begin{itemize}
	\item  Radiative cloud feedback can drive spontaneous atmospheric variability in both temperature and cloud structure under conditions appropriate for BDs and directly imaged EGPs. The variability  mainly comprises cycles in which clouds gradually dissipate after replenishment, and throughout the cycle, the cloud-base altitude, cloud-top altitude and thus the cloud thickness vary over time. The atmospheric vertical temperature profile also evolves along with the cloud cycle. The typical periods of variability are one to tens of hours with typical  amplitude of the variability up to hundreds of Kelvins in effective temperature. This is a novel, very natural mechanism to explain the observed variability on L and T dwarfs.
	\item  The mechanism responsible for the dissipation of clouds is simply the stratification below the cloud base due to cloud radiative heating, which suppresses the rapid supply of deep vapor and causes the net cloud sinks due to particle settling through the cloud base. After the cloud dissipates, atmosphere near the cloud base cools off, which can either cause supersaturation of vapor or lessen the stratification (so that vapor can be mixed above the condensation level by convection), forming new clouds again.  Over the cloud cycle, the cloud top grows to higher altitudes. The responsible mechanism is that, the decreasing outgoing thermal flux due to the rising cloud-top altitude cools the atmosphere above the cloud top, and this cooling tends to drive a large temperature lapse rate just above the cloud top due to the large vertical gradient of cloud mixing ratio near the cloud top, and thus causes the instability to extend slightly higher in altitude. This then  cools the above-cloud temperature profile even more, causing the cloud top to extend to even higher altitude.    
In our model, no artificial forcing such as parameterized convective perturbations are necessary for the variability to occur. It is a totally spontaneous, natural behavior of the system. This differs for instance from the radiative perturbations explored in the 1D model by \cite{robinson2014} which had to be triggered by a perturbation imposed by hand deep in the domain.
	\item  The existence of variability is robust over a wide range of parameter space. However, the detailed evolution of variability is sensitive to model parameters. In general, the variability can be divided into regular or irregular types, the latter associated with the onset of chaos.
	\item  The radiative-cloud-driven variability is appealing to explain the observed flux variability in brown dwarfs and some directly imaged EGPs, especially those evolving irregularly over short timescales. It is also a promising mechanism for cloud breaking, which has been proposed to explain properties of the L to T  dwarf transition.
	\item  A firm prediction is that thick clouds usually correlate to a convective and cooler profile, and thin clouds correlate to a stratified and warmer profile, which is a result of coevolution of clouds and thermal structure. This combination of surface types has been inferred for a few BDs, and retrievals for more variable BDs in the future will help clarify whether this mechanism is prevalent among BDs. 
	
\end{itemize}

We anticipate future endeavor in this direction: first, coupling of radiative cloud feedback into global atmospheric circulation models is essential to truly decipher the observed variability. Second, employing realistic radiative transfer and chemistry is key to understand the spectroscopically resolved variability. Third, coupling with detailed cloud microphysics models will gain significant insights on the underlying mechanisms triggering the L to T transition.

\parskip = \baselineskip
\noindent \emph{Acknowledgement:} We thank Tad Komacek, Vivien Parmentier and Xi Zhang for helpful discussion. We thank the referee for comments that improve the manuscript. This work was supported by NASA Headquarters under the NASA Earth and Space Science Fellowship (Astrophysics) Program  to X.T. and NSF grant AST 1313444 to  A.P.S.

\bibliographystyle{apj}
\bibliography{draft}

\appendix
\section{Vertical Resolution Test}
\label{ch. verticalresolution}
\begin{figure*}      
\epsscale{1.2}      
\plotone{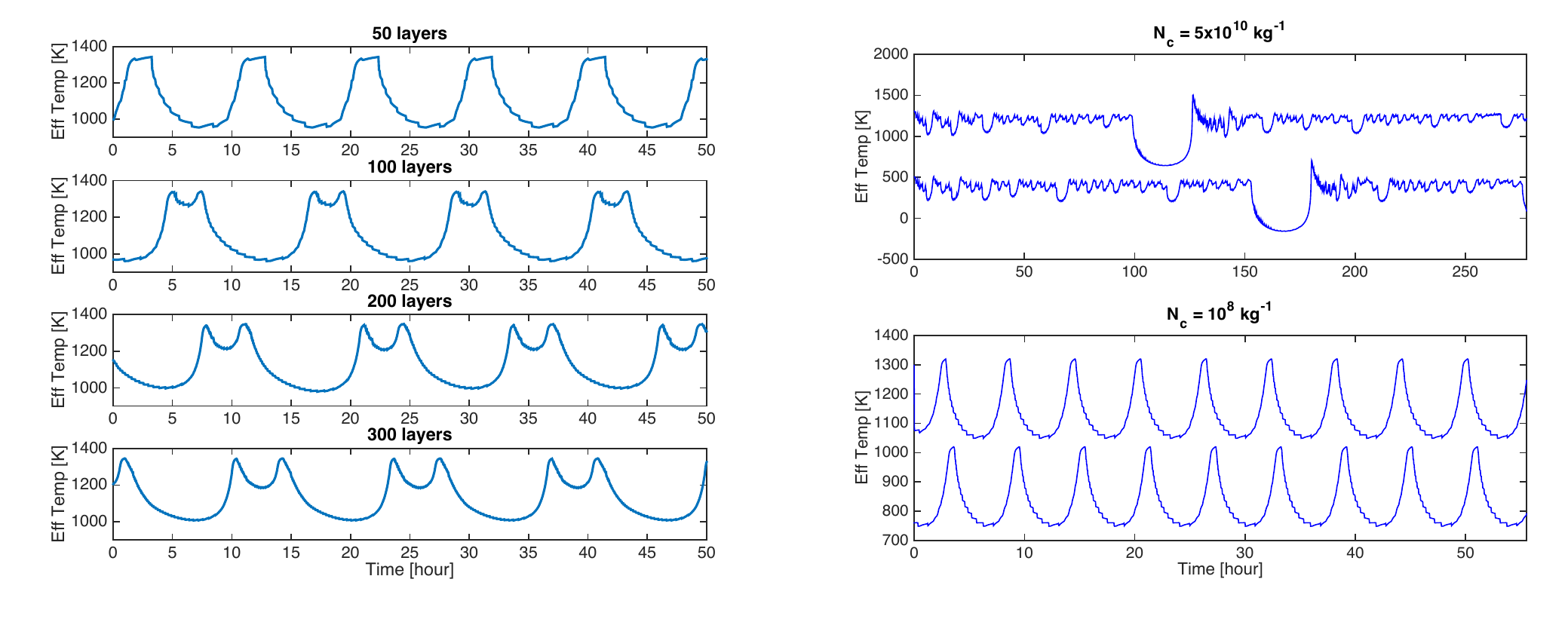}
\caption{\emph{Left:} outgoing thermal flux in terms of effective temperature as a function of time for cases with vertical resolution of 50, 100, 200 and 300 layers. Note that the 100-layer resolution is used in models presented in this work. Results show good agreement for models with 100 layers and more. \emph{Right:} in the upper panel, arbitrarily offset fluxes as a function of time for two cases with $\mathcal{N}_c=5\times10^{10}~ \rm{kg^{-1}}$, in which the initial conditions are differed by only  1\% of cloud mixing ratio in one grid point. In the lower panel, fluxes for two cases with $\mathcal{N}_c= 10^{8}~ \rm{kg^{-1}}$,  in which the first case initially has no cloud and the second case has a thick cloud layer.
}
\label{combine}
\end{figure*}

Here we show vertical resolution test of our model with 50, 100, 200 and 300 vertical layers in the left panels of Figure \ref{combine}, in which the 100-layer resolution is used for models presented in this work. Resolution much higher than 300 layers is too computationally expensive to integrate to a quasi-equilibrium state primarily because of the extremely short diffusion timescale between adjacent model layers. Results show good agreement for models with 100 layers and more. As shown in the sensitivity studies, the detailed behavior of the variability is way more sensitive to physical assumptions of the model than resolutions higher than 100 layers. For the sake of understanding physical mechanisms, the 100-layer resolution is sufficient.

\section{Sensitivity to Initial Conditions}
\label{ch.chaos}

We perform sensitivity studies to  initial conditions for the model with $\mathcal{N}_c=5\times10^{10}~ \rm{kg^{-1}}$ and $\mathcal{N}_c= 10^{8}~ \rm{kg^{-1}}$.  In the right panels of Figure \ref{combine}, the upper panel  shows evolution of fluxes that are offset arbitrarily for two cases, in which the initial conditions are differed by only  1\% of cloud mixing ratio in one grid point. It is easy to see that the evolution of the two models is sensitive to initial conditions. For models with $\mathcal{N}_c= 10^{8}~ \rm{kg^{-1}}$ shown in the lower panel, we test experiments with vastly different initial cloud structures, in which the first case initially has no cloud the second case has a thick cloud layer. One can see that the evolution of fluxes quickly merge to periodic oscillations with almost the same frequency and amplitude, suggesting this model is not sensitive to initial conditions.

\section{Exponential Cloud Distribution}
\label{ch.expcloud}

\begin{figure*}      
\epsscale{0.6}      
\plotone{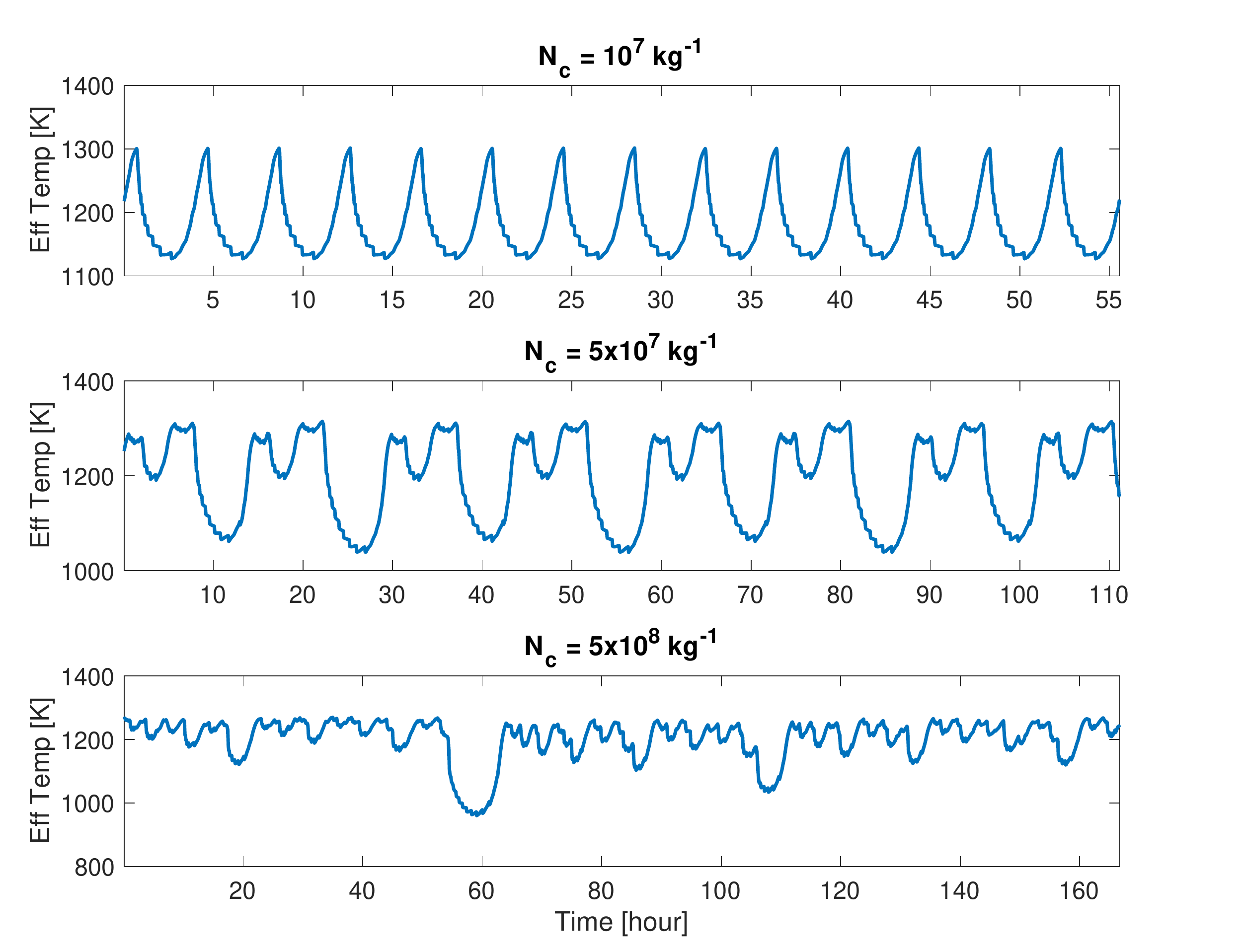}
\caption{Effective temperature as a function of time for models assuming  exponential cloud size distribution with three different cloud number per mass $N_c = 10^7, 5\times 10^7$ and $5\times 10^8 ~\rm{kg^{-1}}$. 
}
\label{expdistribution}
\end{figure*}

As stated in Section \ref{ch.clouddistribution}, we have tested our model assuming an alternative cloud size distribution function -- the exponential distribution described by Eq. (\ref{eq.expcloud}). The variability in terms of effective temperature for models with different cloud number per mass $N_c$ are shown in Figure \ref{expdistribution}. Here we see the same trend as in Section \ref{ch.Nc} wherein smaller cloud number per mass (and hence larger cloud particles) results in regular and higher-frequency oscillations, whereas large cloud number per mass (small cloud particles) exhibits chaotic oscillations. In between, model with mediate $N_c$ shows quasi-regular oscillations. Diagnosis similar to that in Section \ref{nominal} show that the same physical mechanism   is responsible for variability assuming exponential cloud size distribution.

\end{document}